\documentclass[aps,pra,twocolumn,floatfix,superscriptaddress,showpacs]{revtex4}

\usepackage{amssymb}
\usepackage{latexsym}
\usepackage{amsmath}
\usepackage{graphicx}
\usepackage{float}
\usepackage[applemac]{inputenc}
\usepackage[nooneline]{subfigure}

\begin{document}

\title{Different orderings in the narrow-band limit of the extended Hubbard model on the Bethe lattice}

\author{F. Mancini}
\affiliation{Dipartimento di Fisica {\it ``E. R. Caianiello"} \\
Universit\`a degli Studi di Salerno, Via Ponte don Melillo I-84084
Fisciano (SA), Italy} \affiliation{Unit\`a CNISM di Salerno}

\author{F. P. Mancini}
\affiliation{Dipartimento di Fisica {\it ``E. R. Caianiello"} \\
Universit\`a degli Studi di Salerno, Via Ponte don Melillo I-84084
Fisciano (SA), Italy} \affiliation{Laboratorio Regionale SuperMat,
CNR-INFM} \affiliation{I.N.F.N. Sezione di Perugia, Via A.
Pascoli, I-06123 Perugia, Italy}

\begin{abstract}
We present the exact solution of a system of Fermi particles
living on the sites of a Bethe lattice with coordination number
$z$ and interacting through on-site $U$ and nearest-neighbor $V$
interactions. This is a physical realization of the extended
Hubbard model in the atomic limit. Within the Green's function and
equations of motion formalism, we  provide a comprehensive
analysis of the model and we study the phase diagram at finite
temperature in the whole model's parameter space, allowing for the
on-site and nearest-neighbor interactions to be either repulsive
or attractive. We find the existence of critical regions where
charge ordering  ($V>0$) and phase separation ($V<0$) are
observed. This scenario is endorsed  by the study of several
thermodynamic quantities.

\end{abstract}

    \pacs{
      {71.10.Fd Lattice fermion models}
      {71.10-w Theories and models of many-electron systems}
     }

\date{\today}

\maketitle

\section{Introduction}

In recent years, many theoretical as well as experimental
investigations in condensed matter physics have been devoted to
the study of low-dimensional strongly correlated electron systems
where long-range Coulomb interactions play an important role. A
crucial problem is to understand the effects of competing
interactions and the corresponding phase transitions. One of the
seminal models adopted to take into account long-range Coulomb
interactions is the so-called extended Hubbard model (EHM), which,
beside the on-site interaction $U$, contains also a
nearest-neighbor interaction $V$. Although the EHM looks
deceptively simple and in spite of a very intensive study, both
analytical and numerical, there is no  complete exact solution
even in $1D$. These difficulties have led to the investigation of
the so-called atomic limit of the extended Hubbard model (AL-EHM).
According to the conventional definition used in the literature,
the atomic limit stands for the classical limit of the model,
where the hopping matrix elements $t_{ij}$ are set to zero from
the very beginning. The AL-EHM is exactly solvable in one
dimension (see Ref. \cite{mancini08_PRE} and references therein),
as well as when it is defined on the Bethe lattice
\cite{mancini08_1,mancini08_2}. Here we consider the Bethe lattice
as the infinite version of a Cayley tree; thus we are not
concerned with surface effects. The Bethe lattice, beside being an
useful framework for studying the electronic structure of
amorphous and glassy solids, has a relevant role in condensed
matter physics and statistical mechanics. Due to its peculiar
structure, on the Bethe lattice it is possible to exactly solve
several interesting physical problems involving interactions
\cite{baxter}. There are two special properties that make Bethe
lattices particularly suited for theoretical investigations: the
self-similar structure which may lead to recursive solutions and
the absence of closed loops which restricts interference effects
of quantum-mechanical particles in the case of nearest-neighbor
coupling. Furthermore, Bethe and Bethe-like lattices have
attracted a lot of interest because they usually reflect essential
features of systems even when conventional mean-field theories
fail \cite{bethe}. The reason is that such lattices are capable to
take into account correlations which are usually lost in
conventional mean-field calculations. Very recently, the Bethe
lattice has been considered as the underlying lattice to study the
suppression of the paramagnetic metal-insulator transition in the
Hubbard model at half-filling in the presence of nearest-neighbor
and next-nearest-neighbor hoppings when the latter is increased
with respect to the former \cite{NNN1,NNN2}. Furthermore, there
exist in nature hyperbranched polymers with distinct regular
molecular architecture (so-called dendrimers) which can be
conveniently modelized by a Bethe-like Hamiltonian
\cite{dendrimers1,dendrimers2,dendrimers3}.

In this article, we study the AL-EHM on the Bethe lattice by means
of the equations of motion approach \cite{mancini_avella_04}. For
one-dimensional lattices, or more generally for lattices with no
closed loops, classical fermionic and spin systems can be easily
solved by means of the transfer matrix method \cite{baxter}.
However, this method is hardy implementable when more complex
lattices are considered. The Onsager solution for the
two-dimensional Ising model is an emblematic example
\cite{onsager}. We feel that there is the necessity to foster
alternative methods which can be used for a large class of
lattices. In Ref. \cite{mancini_05}  we have shown that the
equations of motion method provides an answer to this exigency. By
using this method, localized fermionic systems and Ising-like
models can be in principle solved for any underlying lattice.
Indeed, one can find a set of eigenenergies and eigenoperators of
the Hamiltonian which closes the hierarchy of the equations of
motion. As a result, one can derive exact analytical expressions
for the relevant Green's functions and correlation functions,
which turn out to depend on a finite set of parameters. The
knowledge of these parameters is essential to obtain a solution of
these models. In a series of articles
\cite{mancini08_PRE,mancini08_1,mancini08_2,mancini_05,mancini_05_2,mancini_naddeo_06,spin32,cmp},
we have analyzed several fermionic and spin systems and we have
developed a self-consistent method which allows us to determine
these parameters for the case of one-dimensional and Bethe
lattices.
Some preliminary results for the AL-EHM on the Bethe lattice were
presented in Refs. \cite{mancini08_1,mancini08_2}, where we
considered only the case of attractive intersite interactions
($V<0$). Upon varying the temperature, we found a region of
negative compressibility, hinting at a transition from a
thermodynamically stable to an unstable phase, characterized by
phase separation. In this paper, we provide a comprehensive and
systematic exact analysis of the AL-EHM on the Bethe lattice with
coordination number $z$ by considering relevant response and
correlation functions as well as thermodynamic quantities. We
consider both the cases $V<0$ and $V>0$ and  cover a wide
range of values of the parameters $n$, $T$ and $U$
 ($n$ is the particle density and $T$ the temperature).
The possibility for the parameters $U$ and $V$ to take positive as
well as negative values - representing effective interaction
couplings taking into account also other interactions (for
instance with phonons) - gives rise to a rich phenomenology and a
variety of phases. In particular, the ratio of the two competing
terms in the model Hamiltonian, i.e., the on-site and the
intersite interactions, determines the different distributions of
the electrons and the accessible phases, which we identified as
charge ordered (CO) or phase separated (PS). Indeed, several
studies of the EHM on regular lattices have pointed at the
presence of states with phase separation
\cite{pha_sep,dongen_95,lin_00,tong_04} and of charge-ordered
phases \cite{lin_00,tong_04,pietig_99,hoang_02,seo_06,pawl_06}.
Furthermore, the introduction of an intersite interaction can
mimic longer-ranged Coulomb interactions needed to describe
effects observed in conducting polymers \cite{baeriswyl_92}. In
this article, we report the phase diagram in the space ($U,n,T$)
showing the existence of critical regions where charge ordering
($V>0$) and phase separation ($V<0$) are observed. These studies
have not been reported in the literature and bring some more
information on the properties of the Hubbard model.

The plan of the paper is as follows. In Sec. \ref{sec_II}, we
review and extend the analysis of the AL-EHM defined on a Bethe
lattice with coordination number $z$ leading to the computation of the Green's
and correlation functions \cite{mancini08_1,mancini08_2} . In Sec. \ref{sec_III}, the attractive
$V$ case is reviewed in detail and the phase diagram is reported
for a wide range of values of the external parameters $n$,
$T/\left| V\right|$, $U/\left| V\right|$. In Sec. \ref{sec_IV}, we
investigate the case of repulsive intersite interactions. The
phase diagram in the space $n$, $T/V$, $U/V$ is derived and a
long-range CO state is observed. The lattice is characterized by a
inhomogeneous distribution of the particles in alternating shells.
Relevant thermodynamic quantities - such as the specific heat,
susceptibility, entropy - are also investigated as functions of
the temperature, on-site potential and particle density. Finally,
Sec. \ref{sec_V}  is devoted to our concluding remarks while the
appendix reports some relevant computational details.

\section{The Hamiltonian and the equations of motion}
\label{sec_II}

The theoretical framework leading to the exact solution of the
AL-EHM defined on the Bethe lattice has been already reported in
Refs. \cite{mancini08_1,mancini08_2}. In this Section we shall
review the analysis and extend it, considering also the breaking
of translational invariance and the addition of an external
magnetic field.

In the extended Hubbard model, a nearest-neighbor interaction $V$
is added to the original Hubbard Hamiltonian, which contains only
an on-site interaction $U$:
\begin{equation}
\begin{split}
\label{ch}
H &=\sum_{\langle \bf ij \rangle} [t_{\bf ij} -\delta
_{\bf ij}\, \mu ]c^\dag (i)c(j)+U\sum_i {n_\uparrow
(i)n_\downarrow (i)} \\
&+\frac{1}{2}\sum_{\langle \bf ij \rangle}
V_{\bf ij} \, n(i)n(j).
\end{split}
\end{equation}
$U$ and $V$ are the strengths of the local and intersite
interactions, respectively; $\mu$ is the chemical potential,
$n(i)=n_{\uparrow }(i)+n_{\downarrow }(i)$ and $D(i)=n_{\uparrow
}(i)n_{\downarrow }(i)=n\left( i\right) \left[ n\left( i\right)
-1\right] /2$ are the particle density and double occupancy
operators, respectively, at site $\bf i$; $t_{\bf ij}$ is the
hopping matrix. As usual, $n_{\sigma }(i)=c_{\sigma }^{\dag
}(i)c_{\sigma }(i)$ with $\sigma =\left\{ {\uparrow ,\downarrow
}\right\} $ where $c_{\sigma }(i)$ ($c_{\sigma }^{\dag }(i))$ is
the fermionic annihilation (creation) operator of an electron of
spin $\sigma $ at site $\bf{i}$, satisfying canonical
anticommutation relations. We use the Heisenberg picture:
$i=\left( {\bf i}, t\right)$, where $\bf{i}$ stands for the
lattice vector $\bf{R}_{i}$. In the extreme narrow-band (atomic)
limit the Hamiltonian \eqref{ch} becomes
\begin{equation}
\label{nbh}
H=-\mu \sum_{\bf i} n(i)+U\sum_{\bf i}
D(i)+\frac{1}{2}\sum_{\langle \bf i j \rangle} V_{\bf ij} \,
n(i)n(j).
\end{equation}
We shall study this model on  a Bethe lattice with coordination
number $z$. For this lattice, the Hamiltonian \eqref{nbh} can be
conveniently rewritten as
\begin{equation}
\label{eq4}
H=-\mu n(0)+UD(0)+\sum_{p=1}^z H^{(p)}.
\end{equation}
$H^{(p)}$ is the Hamiltonian of the $p$-th sub-tree rooted at the
central site $(0)$ and can be written as
\begin{equation}
\label{bethe_H}
H^{(p)}=-\mu \,
n(p)+UD(p)+Vn(0)n(p)+\sum_{m=1}^{z-1} H^{(p,m)}.
\end{equation}
Here $(p)$ ($p=1,\ldots z)$ are the nearest-neighbor sites of
$(0)$, also termed the first shell. $H^{(p,m)}$ describes the
$m$-th sub-tree rooted at the site ($p$); $(p,m)$ ($m=1,\ldots
z-1)$ and $(0)$ are the nearest-neighbors of the site ($p$). The
process may be continued indefinitely.

The equations of motion approach in the context of the composite
operator method \cite{mancini_avella_04} - based on the choice of
a convenient operatorial basis - provides us with the exact
solution of the model. For our purposes, the suitable field
operators are the Hubbard operators, $\xi (i)=[1-n(i)]c(i)$ and
$\eta (i)=n(i)c(i)$, which satisfy the equations of motion:
\begin{equation}
\begin{split}
i\frac{\partial }{\partial t}\xi (i)&=-\mu \xi (i)+zV\xi (i)n^{\alpha }(i) ,\\
i\frac{\partial }{\partial t}\eta (i)&=(U-\mu )\eta (i)+zV\eta
(i)n^{\alpha }(i).
\end{split}
\label{eq4a}
\end{equation}
In the following, for a generic operator $\Phi \left( i\right) $ we shall
use the notation $\Phi ^{\alpha}(i)=\sum\nolimits_{p=1}^{z}\Phi (i,p)/z$, where $ (i,p)$ are
the first nearest-neighbors of the site $i$. The Heisenberg
equations \eqref{eq4a} contain the higher-order nonlocal operators
$\xi (i)n^{\alpha }(i)$ and $\eta (i)n^{\alpha }(i)$. By taking
time derivatives of the latter, higher-order operators are
generated. This process may be continued and an infinite hierarchy
of field operators is created. However, since the number $n(i)$
and the double occupancy $D(i)$ operators satisfy the following
algebra
\begin{equation}
\begin{split}
n^{p}\left( i\right) &=n\left( i\right) +a_{p}D(i), \\
D^{p}\left( i\right) &=D(i), \\
n^{p}\left( i\right) D(i)&=2D(i)+a_{p}D(i),
\end{split}
 \label{al_prop}
\end{equation}
where $p  \geq 1$ and $a_{p}=2^{p}-2$, it is straightforward to
establish the following recursion rule
\cite{mancini_05,mancini_05_2}:
\begin{equation}
\lbrack n^{\alpha }(i)]^{k}=\sum_{m=1}^{2z}A_{m}^{(k)}[n^{\alpha
}(i)]^{m}.
\label{eq5}
\end{equation}
The coefficients $A_{m}^{(k)}$ are rational numbers, satisfying
the relations $\sum_{m=1}^{2z}A_{m}^{(k)}=1$ and
$A_{m}^{(k)}=\delta _{m,k}$ $(k=1,\ldots ,2z)$
\cite{mancini08_PRE,mancini_epjb_05}. The recursion relation (\ref{eq5}) allows
one to close the hierarchy of equations of motion.

\subsection{Eigenoperators and eigenvalues}
\label{II_A}

One may define the composite field operator
\begin{equation}
\label{EHM_2} \psi (i)=\left( \begin{array}{*{20}c}
 {\psi ^{(\xi )}(i)}   \\
 {\psi ^{(\eta )}(i)}   \\
\end{array}  \right),
\end{equation}
where
\begin{equation}
\label{EHM_3}
\psi ^{(\xi )}(i)=\left( {{\begin{array}{*{20}c}
 {\xi (i)}   \\
 {\xi (i)[n^\alpha (i)]}   \\
 \vdots   \\
 {\xi (i)[n^\alpha (i)]^{2z}}   \\
\end{array} }} \right), \;
\psi ^{(\eta )}(i)=\left( {{\begin{array}{*{20}c}
 {\eta (i)}   \\
 {\eta (i)[n^\alpha (i)]}   \\
 \vdots   \\
 {\eta (i)[n^\alpha (i)]^{2z}}   \\
\end{array} }}
\right).
\end{equation}
By using the recursion rule \eqref{eq5}, one can show that the
fields $\psi ^{(\xi )}(i)$ and $\psi ^{(\eta )}(i)$ are
eigenoperators of the Hamiltonian \eqref{eq4}
\cite{mancini08_1,mancini08_2}:
\begin{equation}
\label{EHM_4}
\begin{split}
 i\frac{\partial }{\partial t}\psi ^{(\xi )}(i) &=[\psi ^{(\xi
)}(i),H]=\varepsilon ^{(\xi )}\psi ^{(\xi )}(i),
\\
 i\frac{\partial }{\partial t}\psi ^{(\eta )}(i)&=[\psi ^{(\eta
)}(i),H]=\varepsilon ^{(\eta )}\psi ^{(\eta )}(i),
 \end{split}
\end{equation}
where $\varepsilon ^{(\xi )}$ and $\varepsilon ^{(\eta )}$ are the
$(2z+1)\times (2z+1)$ energy matrices
\begin{widetext}
\begin{equation}
\label{eq8} \varepsilon ^{(\xi )}=\left( {{\begin{array}{*{20}c}
 {-\mu } & {zV}  & 0 \hfill & \cdots  & 0& 0
 & 0  \\
 0  & {-\mu }  & {zV}  & \cdots  & 0  & 0
 & 0  \\
 0  & 0  & {-\mu }  & \cdots  & 0  & 0
& 0  \\
 \vdots  & \vdots  & \vdots  & \cdots  & \vdots
 & \vdots  & \vdots  \\
 0  & 0  & 0  & \cdots  & {-\mu }  & {zV}
 & 0  \\
 0  & 0  & 0  & \cdots  & 0  & {-\mu }
& {zV}  \\
 0  & {zVA_1^{(2z+1)} }  & {zVA_2^{(2z+1)} }  & \cdots
 & {zVA_{2z-2}^{(2z+1)} }  & {zVA_{2z-1}^{(2z+1)} }
 &
{-\mu +zVA_{2z}^{(2z+1)} }  \\
\end{array} }} \right)
\end{equation}
\begin{equation}
\label{eq9}
\varepsilon ^{(\eta )}=\left( {{\begin{array}{*{20}c}
 {U-\mu }  & {zV}  & 0  & \cdots  & 0  & 0
 & 0  \\
 0  & {U-\mu }  & {zV}  & \cdots  & 0  & 0
 & 0  \\
 0  & 0  & {U-\mu }  & \cdots  & 0  & 0
& 0  \\
 \vdots  & \vdots  & \vdots  & \cdots  & \vdots
 & \vdots  & \vdots  \\
 0  & 0  & 0  & \cdots  & {U-\mu }  & {zV}
 & 0  \\
 0  & 0  & 0  & \cdots  & 0  & {U-\mu }
& {zV}  \\
 0  & {zVA_1^{(2z+1)} }  & {zVA_2^{(2z+1)} }  & \cdots
 & {zVA_{2z-2}^{(2z+1)} }  & {zVA_{2z-1}^{(2z+1)} }  &
{U-\mu +zVA_{2z}^{(2z+1)} }  \\
\end{array} }} \right)
\end{equation}
\end{widetext}
whose eigenvalues, $E_m^{(\xi )}$ and $E_m^{(\eta )}$, are given
by:
\begin{equation}
\label{energies}
\begin{split}
 E_m^{(\xi )} &=-\mu +(m-1)V, \\
 E_m^{(\eta )} &=-\mu +U+(m-1)V ,
\end{split}
\end{equation}
with $m=1,\ldots ,2z+1$. The Hamiltonian has now been formally
solved since, for any coordination number of the underlying
Bethe lattice, one has found a closed set of eigenoperators and
eigenenergies. As a result, one may compute observable quantities.
This will be done in the next section by using the formalism of
Green's functions (GF).

\subsection{Retarded Green's functions and correlation functions}
\label{II_B}

The knowledge of a set of eigenoperators and eigenenergies of the
Hamiltonian allows one to find an exact expression for the retarded
Green's function
\begin{equation}
 \label{EHM_8}
 G^{(s)}(t-t')=\theta (t-t')\langle\{\psi
^{(s)}({\bf i},t), {\psi^{(s)}}^\dag
 ({\bf i},t')\}\rangle,
\end{equation}
and, consequently, for  the correlation function
\begin{equation}
\label{EHM_9} C^{(s)}(t-t') = \langle \psi^{(s)}({\bf i},t)
{\psi^{(s)}}^\dag ({\bf i},t')\rangle.
\end{equation}
In the above equations, $s=\xi ,\eta$ and $\langle\cdots \rangle$
denotes the quantum-statistical average over the grand canonical
ensemble. It is not difficult to show that, for fermionic
operators, only the on-site correlations are non-zero.
It is not difficult to show that the
retarded GF and satisfies the equation
\begin{equation}
\label{eqGF}
\left[ {\omega -\varepsilon ^{(s)}} \right]G^{(s)}(\omega )=I^{(s)},
\end{equation}
where $G^{(s)}(\omega )$ is the Fourier transform of $G^{\left( s
\right)}(t-{t}')$ and $I^{(s)}=\langle\{\psi ^{(s)}(i), {\psi
^{(s)}}^\dag (i)\}\rangle$ is the
$(2z+1)\times (2z+1)$ normalization matrix. The solution of Eq.
\eqref{eqGF} is \cite{mancini_avella_04}:
\begin{equation}
\label{EHM_10} G^{(s)}(\omega )=\sum_{m=1}^{2z+1} \frac{\sigma
^{(s,m)}}{\omega -E_m^{(s)} +i\delta }.
\end{equation}
Similarly, the correlation function satisfies the equation
\begin{equation}
\label{eqCF}
C^{(s)}(\omega )=-\left[ {1+\tanh \frac{\beta \omega }{2}} \right] \,
{\rm Im}
\left[ {G^{(s)}(\omega )} \right],
\end{equation}
where  $C^{(s)}(\omega )$ is the Fourier transform of $C^{(s)}(t-{t}')$, whose solution is
\begin{equation}
\label{EHM_11}
 C^{(s)}(\omega )=\pi \sum_{m=1}^{2z+1} \sigma
^{(s,m)} \, T_m^{(s)} \, \delta (\omega -E_m^{(s)} ).
\end{equation}
In the above equations, $T_m^{(s)} =1+\tanh \big( \beta E_m^{(s)}/2 \big)$, $\beta
=1/k_B T$ and the $E_{m}^{(s)}$ are given in Eq. \eqref{energies}.
The spectral density matrices $\sigma _{ab}^{(s,n)}$ can be
computed by means of the formula \cite{mancini_avella_04}:
\begin{equation}
\sigma _{ab}^{(s,n)}=\Omega _{an}^{(s)}\sum_{c=1}^{2z+1}\left[ {
\Omega _{nc}^{(s)}}\right] ^{-1}I_{cb}^{(s)}.  \label{eq14}
\end{equation}
In Eq. \eqref{eq14}, $\Omega ^{(s)}$ is the $(2z+1)\times (2z+1)$
matrix whose columns are the eigenvectors of the energy matrix
$\varepsilon ^{(s)}$.  $I_{a,b}^{(s)}$ are instead the elements of
the normalization matrix $I^{(s)}$. Calculations show that $\Omega ^{(\xi
)}=\Omega ^{(\eta )}=\Omega $, with the matrix $\Omega $ given by
\begin{equation}
\label{eq34}
\Omega _{p,k} =\left\{ {{\begin{array}{*{20}c}
 1 \\
 0 \\
 {\left( {\frac{z}{k-1}} \right)^{2z+1-p}} \\
\end{array} }} \right.
\quad \quad
{\begin{array}{*{20}c}
 {k=1,\;p=1} ,\\
 {k=1,\;p\ne 1}, \\
 {k\ne 1} .\\
\end{array} }
\end{equation}
The matrix elements of the normalization matrices $I^{(s)}$ have
the expressions
\begin{equation}
\label{eq341}
 I_{n,m}^{(\xi )}
 =\kappa ^{(n+m-2)}-\lambda ^{(n+m-2)},
\quad
I_{n,m}^{(\eta )}
=\lambda ^{(n+m-2)},
\end{equation}
where the correlators $\kappa ^{(p)}$ and $\lambda ^{(p)}$ are
defined as
\begin{equation}
\label{eq342}
 \kappa ^{(p)}
 =\langle [n^\alpha (i)]^p \rangle,
 \quad \lambda ^{(p)}
 =\frac{1}{2} \langle n(i)[n^\alpha (i)]^p \rangle .
\end{equation}
By exploiting the recursion relation \eqref{eq5}, it is not
difficult to show that also $\kappa ^{(p)}$ and $\lambda ^{(p)}$
obey similar recursion relations
\begin{equation}
\label{eq33}
 \kappa ^{(p)}
 =\sum_{m=1}^{2z} A_m^{(p)} \kappa ^{(m)},
 \quad
 \lambda ^{(p)}
 =\sum_{m=1}^{2z} A_m^{(p)} \lambda ^{(m)},
\end{equation}
limiting their computation to the first $2z$ correlators
\cite{mancini08_1}. At this stage, the knowledge of the GFs and of
the CFs is not yet achieved since they depend on the $\{ \sigma
^{(s,m)} \}$ which, in turn, depend on the normalization matrix
elements: there are $2z$ parameters to determine. To find
these parameters, we shall exploit the Pauli principle and impose
pertinent boundary conditions for obtaining a set of self-consistent
equations.

One first chooses an arbitrary site, say ${\bf i}$,
then splits the Hamiltonian \eqref{eq4} in the sum of two
terms: $H=H_0^{(i)} +H_I^{(i)}$, where
\begin{equation}
\label{EHM_17}
\begin{split}
H_0^{(i)} &=-\mu [n(i)+zn^\alpha (i)]+U[D(i)+zD^\alpha
(i)]\\
&+\sum_{p=1}^z \sum_{m=1}^{z-1} H^{(p,m)} ,\\
 H_I^{(i)} &=zVn(i)n^\alpha (i),
 \end{split}
\end{equation}
and introduce the $H_0^{(i)}$-representation: the statistical
average of any operator $O$ can be expressed as
\begin{equation}
\label{EHM_18}
 \langle O\rangle =\frac{\langle Oe^{-\beta H_I^{(i)}
}\rangle _{0,\bf i}} {\langle e^{-\beta H_I^{(i)} }\rangle_{0, \bf
i} }.
\end{equation}
The symbol $\langle \cdots \rangle_{0, \bf i}$ stands for the
thermal average with respect to the reduced Hamiltonian
$H_0^{(i)}$: i.e., $\langle \cdots \rangle_{0, \bf i} =Tr\{\cdots
e^{-\beta H_0^{(i)}}\}/Tr\{e^{-\beta H_0^{(i)}}\}$. Equation
\eqref{EHM_18} allows us to express the thermal averages with
respect to the complete Hamiltonian $H$ in terms of thermal
averages with respect to the reduced Hamiltonian $H_0$, which
describes a system where the original lattice has been reduced to
the site ${\bf i}$ and to $z$ unconnected sublattices. As a
consequence, in the $H_0$-representation, correlation functions
connecting sites belonging to disconnected sublattices can be
decoupled. Let us consider the correlation functions
\begin{equation}
\label{eq24}
 C_{1,k}^{(s)} =\langle {s(i)s^\dag (i)\left[
{n^\alpha \left( i\right)} \right]^{k-1}}\rangle ,
\end{equation}
where $s=\xi, \eta$ and $k=1,...,2z+1$. By means of Eq.
\eqref{EHM_18}, they can be written as:
\begin{equation}
\label{eq25}
 C_{1,k}^{(s)} =\frac{\langle s(i)s^\dag(i)\left[
n^{\alpha} (i)\right]^{k-1} e^{-\beta H_I }\rangle_{0,\bf i}}{\langle
e^{-\beta H_I }\rangle_{0,\bf i}}.
\end{equation}
The Pauli principle leads to the following algebraic relations
\begin{equation}
\label{eq26}
\begin{split}
 \xi^\dag (i)n(i)&=0, \\
 \xi ^\dag (i)D(i)&=0,
\end{split} \quad \quad \quad
\begin{split}
\eta ^\dag (i)n(i)&=\eta ^\dag (i), \\
 \eta ^\dag (i)D(i)&=0,
\end{split}
\end{equation}
from which one has $\xi ^\dag (i)\,e^{-\beta H_I }=\xi ^\dag (i)$,
and $\eta ^\dag (i)\,e^{-\beta H_I }=\eta ^\dag (i)\,e^{-z\beta
Vn^\alpha (i)}$. In the $H_0$-representation,
the correlation functions can be rewritten as:
\begin{equation}
\label{eq27}
\begin{split}
 C_{1,k}^{(\xi )} &=\frac{\langle {\xi (i)\xi ^\dag
(i)} \rangle_{0,\bf i} \langle {[n^\alpha (i)]^{k-1}} \rangle _{0,\bf i}
}{\langle {e^{-\beta H_I }} \rangle _{0,\bf i} },\\
C_{1,k}^{(\eta )} &=\frac{\langle {\eta (i)\eta ^\dag (i)} \rangle
_{0,\bf i} \langle {[n^\alpha (i)]^{k-1}e^{-z\beta Vn^\alpha (i)}} \rangle
_{0,\bf i} }{\langle {e^{-\beta H_I }} \rangle_{0,\bf i}}.
\end{split}
\end{equation}
In the $H_0$-representation, the Hubbard operators obey to simple
equations of motion: $[\xi (i),H_0 ]=-\mu \,\xi (i)$ and $[\eta
(i),H_0 ]=-(\mu -U)\,\eta (i)$. Thus, it is easy to show that the
equal time CF's can be expressed as:
\begin{equation}
\label{eq28}
\begin{split}
 \langle {\xi (i)\xi ^\dag (i)} \rangle _{0,\bf i} &=\frac{1}{1+2e^{\beta
\mu }+e^{\beta (2\mu -U)}}=1-B_1 +B_2 ,\\
 \langle {\eta (i)\eta ^\dag (i)} \rangle _{0,\bf i} &=\frac{e^{\beta \mu
}}{1+2e^{\beta \mu }+e^{\beta (2\mu -U)}}=\frac{1}{2}\left(B_1 -2B_2 \right),
 \end{split}
\end{equation}
where:
\begin{equation}
\label{eq29}
\begin{split}
 B_1 &=\langle n(i)\rangle_{0,\bf i}=\frac{2 (f +g)}{1+2f +g }, \\
 B_2 &=\langle D(i)\rangle_{0,\bf i} =\frac{g }{1+2f+g }.
 \end{split}
\end{equation}
In Eq. \eqref{eq29}, we have defined $f=e^{\beta \mu}$,
$g=e^{\beta (2\mu-U)}$, and we  have used the identities
\begin{equation}
\label{eq30}
\begin{split}
&\xi _\sigma \xi _\sigma ^\dag +\eta _\sigma \eta _\sigma ^\dag =1-n_\sigma ,
\\
&\eta _\sigma \eta _\sigma ^\dag =n_\sigma -n_\uparrow n_\downarrow .
 \end{split}
\end{equation}
Upon inserting Eqs. \eqref{eq28} into Eqs. \eqref{eq27} and by
taking $k=1$, one finds:
\begin{equation}
\label{eq31}
\begin{split}
C_{1,1}^{(\xi )} &=\frac{1-B_1 +B_2}{\langle e^{-\beta H_I}
\rangle_{0,\bf i}} ,
\\
C_{1,1}^{(\eta )} &=\frac{\left( {B_1 -2B_2 } \right)\langle
e^{-z\beta V n^\alpha (0)}\rangle_{0,\bf i}}{2 \langle e^{-\beta
H_I}\rangle _{0,\bf i}}.
\end{split}
\end{equation}
It is not difficult to show that the averages in the above equations can be
expressed as:
\begin{equation}
\label{eq32}
\begin{split}
&\langle e^{-\beta H_I^{(i)} }\rangle_{0, \bf i} = 1+B_1
\left( F_i^z -1\right) +B_2 \left( 1-2
F_i^z+G_i^z \right),
\\
& \langle {e^{-z\beta Vn^\alpha (0)}} \rangle _{0,\bf i}
 =F_i^z,
 \end{split}
\end{equation}
where
\begin{equation}
F_i = 1+ a X_i +a^2 Y_i , \quad G_i = 1+ d X_i +d^2 Y_i .
\end{equation}
In the above equations we have defined $a= K-1$, $d=K^2-1$, with
$K=e^{-\beta V}$; $i_p$ ($p=1,\ldots,z $) is an arbitrary
neighboring site of $i$. $X_i$ and $Y_i$ are two parameters
defined as:
\begin{equation}
\begin{split}
X_i &= \langle n^{\alpha}(i)\rangle_{0,\bf i} =\frac{1}{z}\sum_{p=1}^z
\langle n(i_p )\rangle_{0,\bf i} ,
\\
Y_i &= \langle
D^{\alpha}(i) \rangle_{0,\bf i} = \frac{1}{z}\sum_{p=1}^z \langle D(i_p
)\rangle_{0,\bf i} .
\end{split}
\end{equation}
$X_i$ and $Y_i$ are parameters of seminal importance since all
correlators and fundamental properties of the system under study
can be expressed in terms of them. Relevant physical quantities,
such as the mean value of the particle density and doubly
occupancy, and the charge correlators $\kappa^{(k)}$ and
$\lambda^{(k)}$  \eqref{eq342} can be easily computed. After
lengthy but straightforward calculations, one finds:
\begin{equation}
\label{eq_nD_1}
\begin{split}
 \langle {n(i)}\rangle &=\frac{2f\,F_i ^z+2g\,G_i ^z}{1+2f\,F_i ^z+ g\,G_i
 ^z},
\\
 \langle {D(i)} \rangle &=\frac{g\,G_i ^z}{1+2f\,F_i ^z+g\,G_i
^z},
\end{split}
\end{equation}
and
\begin{equation}
\label{eq_nD_2}
\begin{split}
\langle {n(i_p )} \rangle &=\frac{1}{1+2f\,F_i ^z+g\,G_i ^z}
\left[ 2f \,K \left( {X_i +2aY_i}
\right) F_i ^{z-1}  \right.
\\
 &+  \left.  g\,K^2 \left( {X_i +2d\,Y_i } \right)G_i
^{z-1} +X_i  \right], \\
 \langle {D(i_p )} \rangle
&=\frac{Y_i \left(1+2f K^2F_i ^{z-1}+g\,K^4 G_i
^{z-1}\right)}{1+2f\,F_i ^z+g\,G_i ^z}.
 \end{split}
\end{equation}
The parameters $X_i$ and $Y_i$ will be fixed by using  boundary
conditions, which will be different according to the sign of the
intersite potential $V$. Therefore, we shall consider separately
the cases $V<0$ and $V>0$. For both cases we shall study, in the
next sections,  relevant thermodynamic quantities and response
functions, such as the internal energy, the  specific heat, the
charge and spin susceptibilities and the entropy. The internal
energy $E$  can be computed as the thermal average of the
Hamiltonian \eqref{nbh} and it is given by $E = U D + z V
\lambda^{(1)}$. The specific heat is then directly given by $C =
dE/dT$. The charge susceptibility $\chi_c$ can be computed by
means of thermodynamics through the formula
\begin{equation}
\label{EHM_35}
 \chi_c = N n^2 + \frac{1}{\beta }\frac{\partial
n}{\partial \mu } .
\end{equation}
In the above equation,  $N$ is the number of sites and $n=\sum_i
\langle n(i) \rangle /N$ is the particle number per site. The spin
magnetic susceptibility $\chi_s$ can be computed by introducing an
external magnetic field $h$, taking the derivative of the
magnetization $m = \langle n_\uparrow (i) - n_\downarrow
(i)\rangle$ with respect to $h$ and letting $h$ going to zero:
\begin{equation}
\chi_s= \left( \frac{ \partial m}{\partial h} \right)_{h = 0}.
\label{spin_susc}
\end{equation}
The addition of a homogeneous magnetic field does not dramatically
modify the framework of calculation given in this section, once
one has taken into account the breakdown of the spin rotational
invariance. Some details of the calculations are given in the
appendix.

\section{Attractive intersite potential}
\label{sec_III}

\begin{figure}[t]
\centering\includegraphics[scale=0.225]{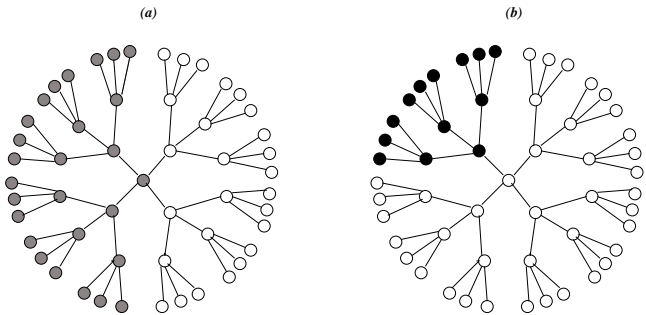}
\caption{\label{fig1} Distribution of the particles in the Bethe
lattice at $n=0.5$ and $T=0$: ($a$) $V<0$ and $U \gtrapprox U_{PS}$;
($b$) $V<0$ and $U \lessapprox U_{PS}$. White, grey and black circles
denote empty, arbitrary spin singly occupied and double occupied
sites, respectively.}
\end{figure}

In this Section, we review  the case of attractive intersite
potential. For $V<0$, the AL-EHM on the Bethe lattice exhibits a
phase separation at low temperatures \cite{mancini08_2}. This
phenomenon is characterized by a  macroscopically inhomogeneous
ground state where different spatial regions have different
average particle densities.  As it has been evidenced in Ref.
\cite{mancini08_2}, there exists a critical temperature $T_c$
below which the system loses translational invariance and phase
separation occurs. Moreover, for $T<T_c$ two types of phase
separated configurations are possible according to the strength of
the on-site potential $U$: clusters of singly occupied sites or
clusters of doubly occupied sites. This is qualitatively
illustrated in Fig. \ref{fig1}, where we report just one possible
configuration for $n=0.5$ and $T=0$. There is a critical value of
the on-site potential $U_{PS}$, depending on the coordination number
$z$ and on filling $n$, separating the two types of phase separated configurations.
In the high temperature regime, one may safely assume that the
system is in the translational invariant phase. Thus, one
requires: $ \langle n(i)\rangle= \langle n^\alpha (i)\rangle$ and
$\langle D(i)\rangle= \langle D^\alpha (i)\rangle$, $\forall i$.
As a consequence, from Eqs. \eqref{eq_nD_1} and \eqref{eq_nD_2},
we obtain two equations allowing us to determine $X_i$ and $Y_i$
as functions of the chemical potential $\mu$:
\begin{subequations}
\begin{equation}
\label{eq_for_X_Y_1} X =2 f (1-X-d Y)F^{z-1}+ g [2+(d-1)X-2d
Y]G^{z-1},
\end{equation}
\begin{equation}
\label{eq_for_X_Y_2} Y =g  [1+d X-(2d+1)Y]G^{z-1} -2f K^{2}Y
F^{z-1}.
\end{equation}
\label{eq_for_X_Y}
\end{subequations}
In the above equations,  we dropped the index $i$ because all the
sites are equivalent. Since experimentally, by varying the doping,
it is possible to tune the density in a controlled way, here we
shall  fix the particle density $n=\langle n(i)\rangle$; the
chemical potential will be determined by the system itself,
according to the values of the external parameters, by means of
the following equation:
\begin{equation}
n=\frac{\left( X-2Y\right) F +2YG}{ 1-X+Y +\left(
X-2Y\right) F +Y G}.
\label{chem1}
\end{equation}
Equations \eqref{eq_for_X_Y} and \eqref{chem1} constitute a system
of coupled equations allowing one to ascertain the three
parameters $\mu$, $X$ and $Y$ in terms of the external parameters
of the model $n$, $U$, $V$, $T$, and $z$. Once these quantities
are known, all the properties of the model can be computed. As an
example, the double occupancy $D=\langle D(i)\rangle$ and the
nearest-neighbor charge correlation function $\lambda
^{(1)}=\langle {n(i)n^{\alpha }(i)]}\rangle $/2, in terms of $X$
and $Y$, are given by:
\begin{equation}
\begin{split}
D &=\frac{Y G }{ 1-X+Y+\left( X-2Y\right) F +Y G },
 \\
\lambda ^{(1)} &=\frac{K \left( X+2aY\right) \left( X-2Y\right)
+2K^{2}Y\left( X+2dY\right) }{2\left( 1-X+Y\right) +2\left(
X-2Y\right) F +2Y G }.
\end{split}
\label{quantities2}
\end{equation}


\subsection{The solution at half filling}

As it has been evidenced in Ref. \cite{mancini08_1}, of particular
interest is the case of half filling due to the possibility to
analytically reveal the spontaneous breakdown of the particle-hole
symmetry. Because of this invariance property of the Hamiltonian
\eqref{eq4}, at $n=1$ the chemical potential does not depend on
the temperature and takes the constant value $\mu=U/2+zV$. Upon
substituting this value of $\mu$ in Eqs. \eqref{eq_for_X_Y}, one
finds that these equations admit the following solution
\begin{subequations}
\label{eq40}
\begin{equation}
\label{eq40a}
 X =1-dY ,
\end{equation}
with $Y$ determined by the equation
\begin{equation}
Y \left( 1+K^2 \right)+2e^{\beta U/2}KY  \left( 1-a^2
Y\right)^{z-1}-1=0.
\end{equation}
\end{subequations}
It can be shown that this solution gives
\begin{equation}
\label{eq41}
\begin{split}
 n&=1,\\
  D&=\frac{1}{2+2 e^{\beta U/2}(1-a^2Y )^z},
 \\
  \lambda ^{(1)}& = \frac{1}{2}+d Y D.
  \end{split}
\end{equation}
That is, solution \eqref{eq40} satisfies the particle-hole symmetry.
On the other hand, if one perturbs the solution \eqref{eq40a}, by
setting for example $X =1-dY +\delta$, it is straightforward to
show that a particle-hole symmetry breaking solution exists
for temperatures lower than a critical temperature $T_c$, determined by:
\begin{equation}
\label{eq43}
\begin{split}
&2K^{\frac{U/\left| V
\right|+2}{2}}[z+K(z-2)]^{z-1} \\
&+(K+1)^{z-1}(z-1)^{z-1}[z-K^2(z-2)]=0
.
\end{split}
\end{equation}
\begin{figure}[t]
\vspace{-5mm}
 \centering
 \includegraphics[scale=0.2251]{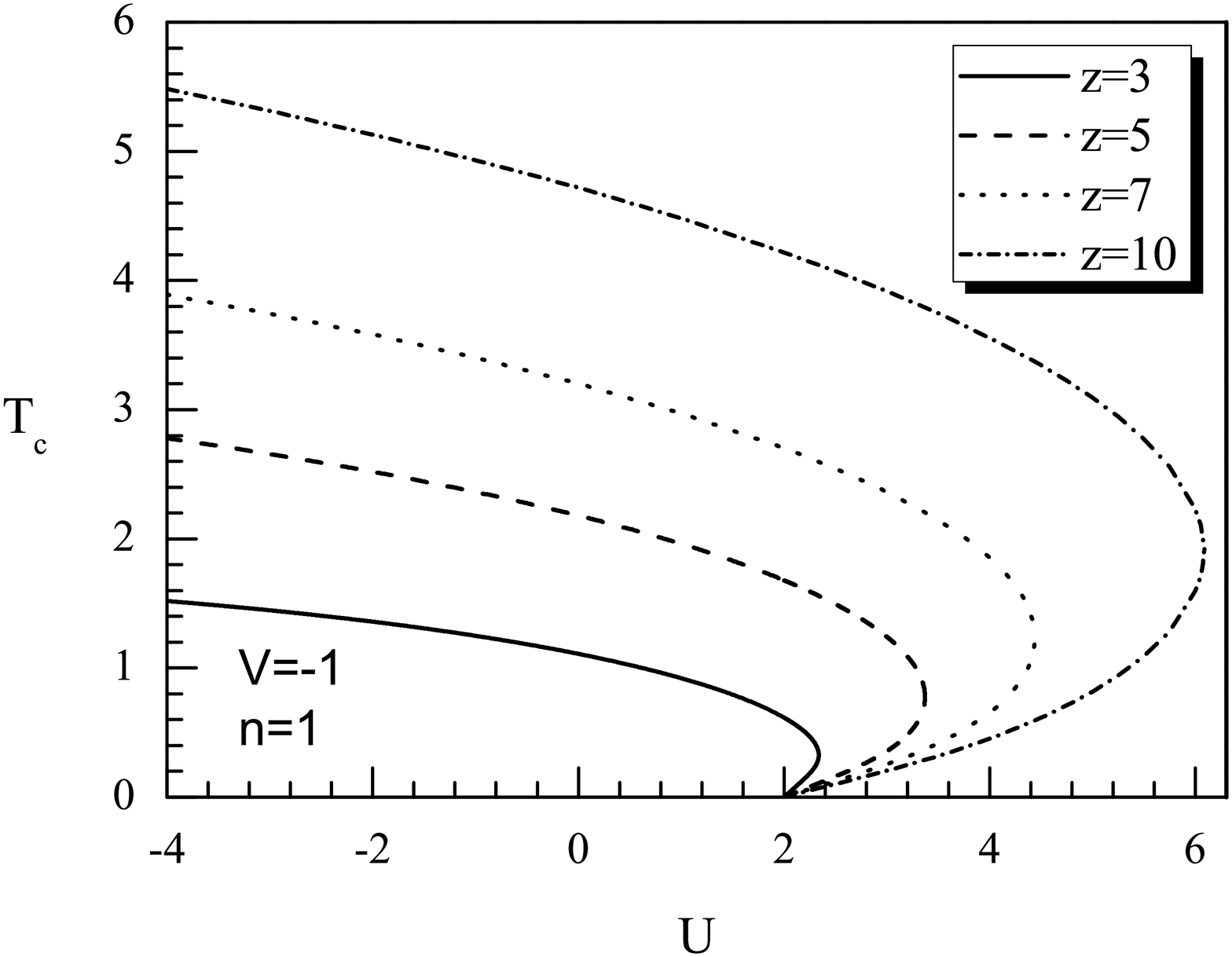}
\caption{\label{figTc_vs_z} The transition temperature $T_c$ as a
function of the on-site potential $U$ for $V=-1$, $n=1$ and
different values of the coordination number $z$.}
\end{figure}
Numerical calculations show that Eq. (\ref{eq43}) admits a
solution only for attractive intersite interactions. It is
interesting to consider Eq. \eqref{eq43} in two extremal limits,
namely $U/\left| V \right|\to -\infty $ and $T_c \to 0$. In the
former limit, corresponding to the case of vanishing intersite
potential, one finds the same critical temperature exhibited by a
system of spinless fermions living on the sites of a Bethe lattice
\cite{mancini08_1}:
\begin{equation}
\label{eq44} \frac{k_B T_c }{\left| V \right|}= 2 \left[ \ln
\left( {\frac{z}{z-2}} \right) \right]^{-1}
\end{equation}
Since the spinless fermion model can be mapped into the spin-1/2
Ising model, our result for the critical temperature agrees with
the one previously found in the literature \cite{baxter,mancini_naddeo_06}.
In the second limit, Eq. \eqref{eq43} becomes
\begin{equation}
\label{eq45}
\left( {U/2-\left| V \right|} \right)=k_B T_c
\ln \left[ {\frac{(z-1)^{z-1}}{2(z-2)^{z-2}}} \right].
\end{equation}
That is,  $U =2\vert V\vert$ is the critical value of the on-site
potential at which a quantum phase transition occurs. The results
obtained from Eq. \eqref{eq43} are displayed in Fig.
\ref{figTc_vs_z}. One observes that, at fixed coordination number,
by increasing $U$ from large negative values, the critical
temperature decreases, and the lower the coordination number, the
lower the critical temperature. An interesting feature of the
phase diagram is that for $U>2\vert V\vert$, the critical
temperature exhibits a reentrant behavior. The width of the
reentrance increases with $z$, and the turning point is at a
critical value $U_c$, which depends linearly on the coordination
number via the law: $U_c / \vert V \vert= a+b z$ (for $z \ge 3$),
where  $a \approx 0.69$ and $b \approx 0.54$. For $z=3$ one has
$U_c \approx 2.3 \vert V \vert$.

\subsection{Phase diagram and local properties}

In this Subsection, we review the case of arbitrary filling
analyzed in Ref. \cite{mancini08_2}. By solving the set of
equations \eqref{eq_for_X_Y} and \eqref{chem1}, it is possible to
derive the phase diagram and various local properties in terms of
the external parameters $n$, $U$, and $T$, taking $\vert V \vert$
as the unit of energy.
\begin{figure}[t]
 \centering
 \subfigure[]
   {\includegraphics[scale=0.5]{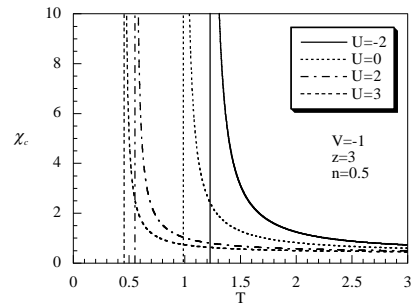}}
   \subfigure[]
{\includegraphics[scale=0.5]{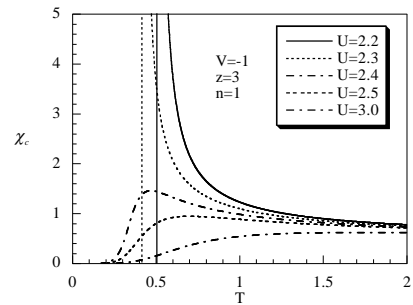}}
\caption{\label{fig79} The charge susceptibility $\chi_c$  for
$V=-1$ and $z=3$ as a function of the temperature $T$
for several values of $U$ and (a) $n=0.5$;  (b) $n=1$. }
 \end{figure}
An important quantity useful for studying the critical behavior of
the system is the susceptibility. In Figs. \ref{fig79}, we plot
the charge susceptibility as a function of the temperature for
$n=0.5$ and $n=1$, and for several values of the on-site
potential. From Fig. \ref{fig79}a, one observes that there is a
critical temperature - depending on $n$ and $U$ - at which the
charge susceptibility $\chi_c$ diverges,  for both attractive and
repulsive $U$. Below $T_c$, $\chi_c$ becomes negative (not
reported in the graphs). As a result, one immediately infers that
there exists a  region  where the system is thermodynamically
unstable.  At fixed $U$, the instability is observed in a particle
density region $\Delta n=n_{1}\leq n\leq n_{2}$, whose width
varies with the temperature, and vanishes for $T> T_c$
\cite{mancini08_2}. For values of the particle density less than
half filling, the divergence of the charge susceptibility is
observed for all values of $U$. As it is shown in  Fig.
\ref{fig79}b, for $n=1$ and for $U> U_c$, no phase transition is
observed: $\chi_c$ is well defined for all values of $T$ and
vanishes in the limit $T \to 0$. It can be shown that, in the
opposite limit $T\to \infty$, $\chi_c$ tends to a constant value
which does not depend on $U$ but only on $n$ according to the law
\begin{equation*}
\lim_{T\to \infty } \chi _c =n\left( 1-\frac{n}{2}\right).
\end{equation*}
%
\begin{figure}[t]
 \centering
 \subfigure[]
   {\includegraphics[scale=0.5]{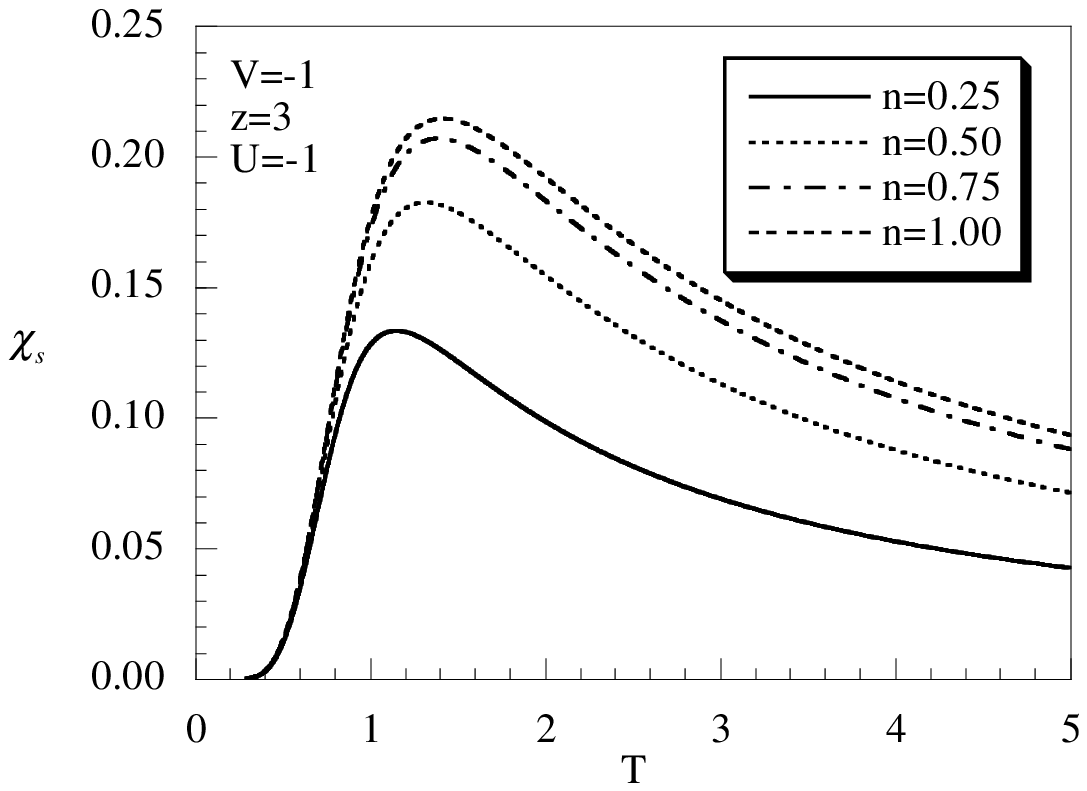}}
 \subfigure[]
   {\includegraphics[scale=0.5]{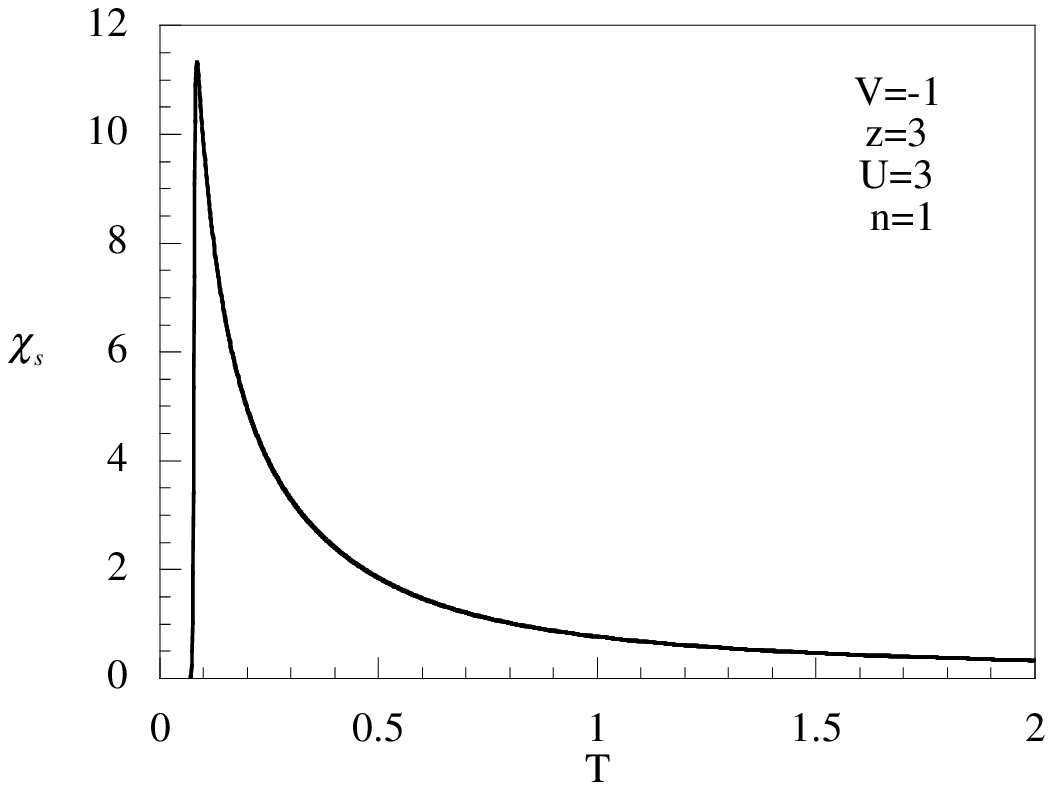}}
\caption{\label{fig80} The spin magnetic susceptibility $\chi_s$
as a function of the temperature for $V=-1$ and $z=3$,
and for (a) $U=-1$ and $n=0.25$, 0.5, 0.75, 1; (b) $U=3$ and
$n=1$.}
 \end{figure}
In Figs. \ref{fig80}, we plot the spin susceptibility $\chi_s$ as
a function of the temperature. For attractive intersite
interaction, at $T=T_c$ one does not observe substantial changes
in the behavior of $\chi_s$. For $T>T_c$, the system is in the
high-temperature regime and $\chi_s$ follows a Curie law. As it is
evident from Fig. \ref{fig45}b, the only low-temperature region
accessible to our investigation is when $n=1$ and $U>U_c$. In Fig.
\ref{fig80}b, we show the spin magnetic susceptibility  $\chi_s$
as a function of $T$ for $n=1$ and $U=3$. For  $U>U_c$, $\chi_s$
is rather insensitive to the value of $U$, and a large peak is
observed at low temperatures. The behavior of the spin
susceptibility for repulsive intersite interactions is more
relevant, as we will show in the next section,  since it signals
the occurrence of a CO phase.

In Fig. \ref{fig45}a, we report the phase diagram in the 3D space
($U,n,T$). The critical temperature $T_{c}$ and the width of the
instability region $\Delta n$ increase by decreasing $U$: an
attractive on-site potential will favor phase separation and in
particular the clustering of doubly occupied sites. As a function
of the particle density, the critical temperature shows a
lobe-like behavior: it increases by increasing $n$ up to half
filling, where it has a maximum; further augmenting $n$, it
decreases vanishing at $n=2$. A different behavior is observed
when $U>2\left| V\right|$: the lobe splits in two and $T_c$
increases with $n$ up to quarter filling, then decreases with a
minimum at half filling. The critical temperature at $n=1$ is
finite only for $U<U_c$. From Fig. \ref{fig45}b, one immediately
notices that, for larger values of the on-site
potential, there is no transition at half filling, as it was noticed in
Ref. \cite{mancini08_2}.
\begin{figure}[t]
 \centering
 \subfigure[]
   {\includegraphics[scale=0.6]{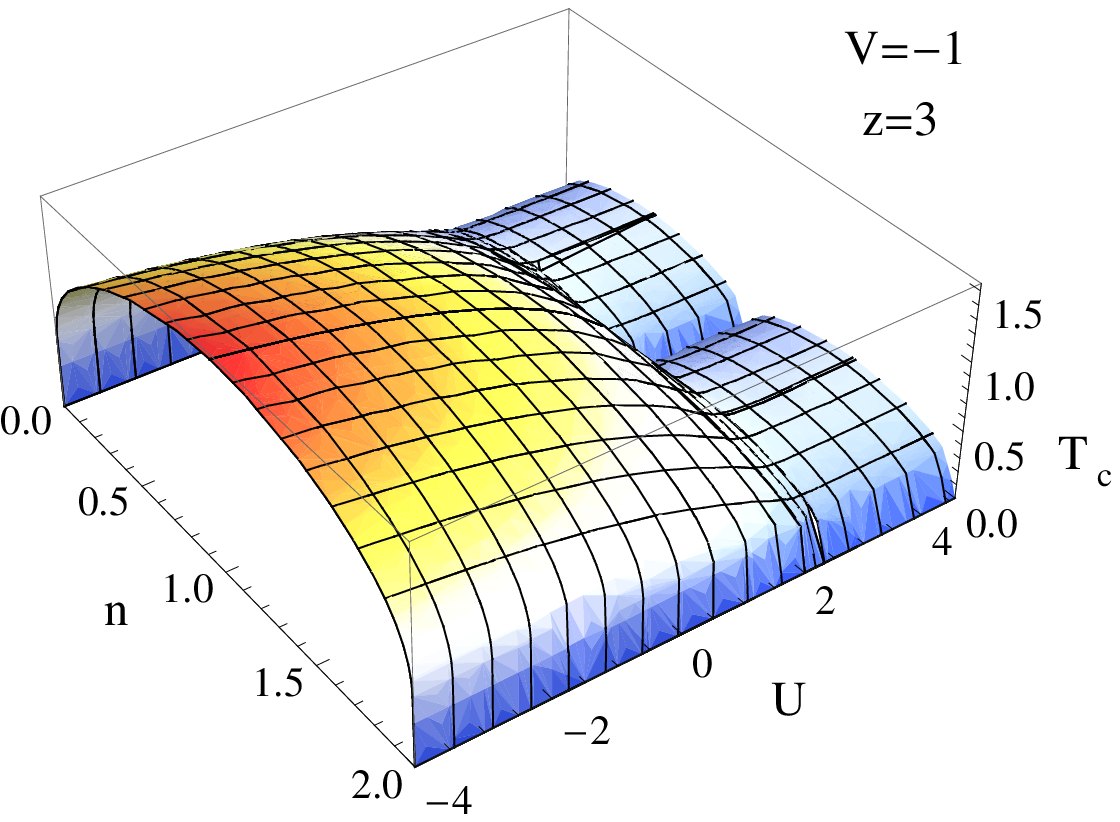}}
 \subfigure[]
   {\includegraphics[scale=0.52]{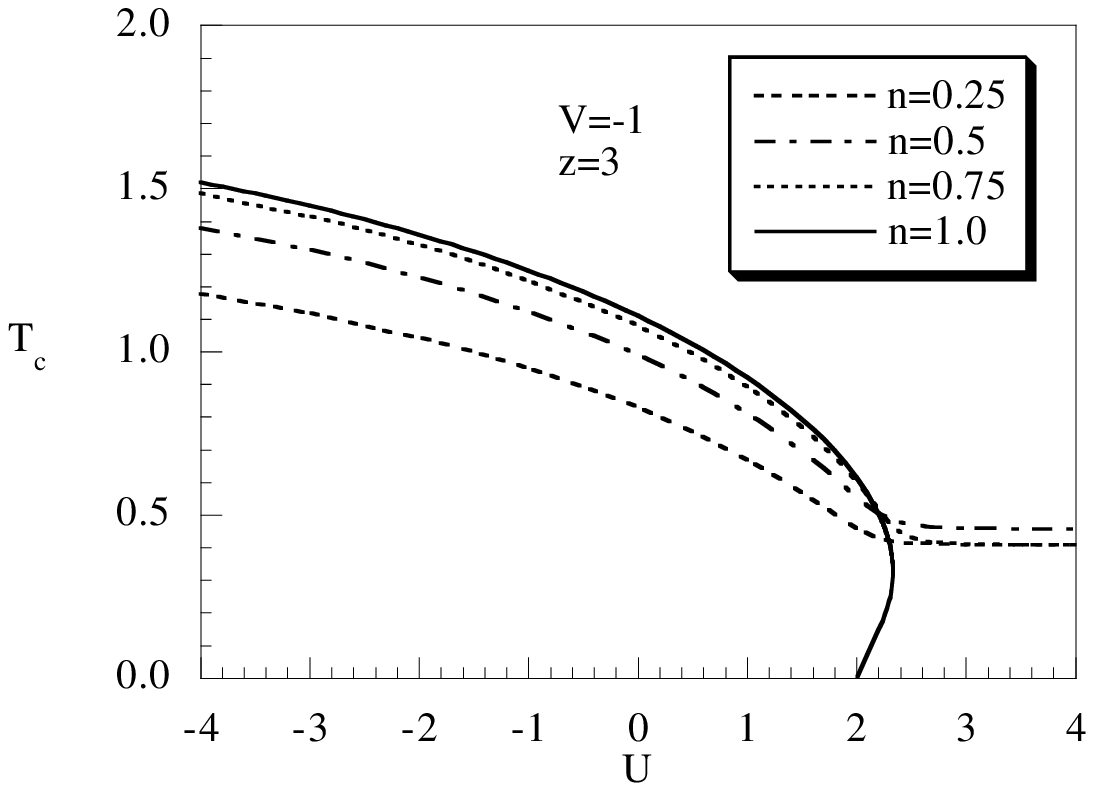}}
\caption{\label{fig45}(a) (Color online) The critical temperature
$T_{c}$ as a function of the particle density $n$ and of the
on-site potential $U$ for $V=-1$ and $z=3$. (b) The critical
temperature $T_{c}$ as a function of the on-site potential $U$ for
$V=-1$, $z=3$ and for $n=0.25$, 0.5, 0.75, 1.}
 \end{figure}

\begin{figure}[t]
 \centering
\includegraphics[scale=0.45]{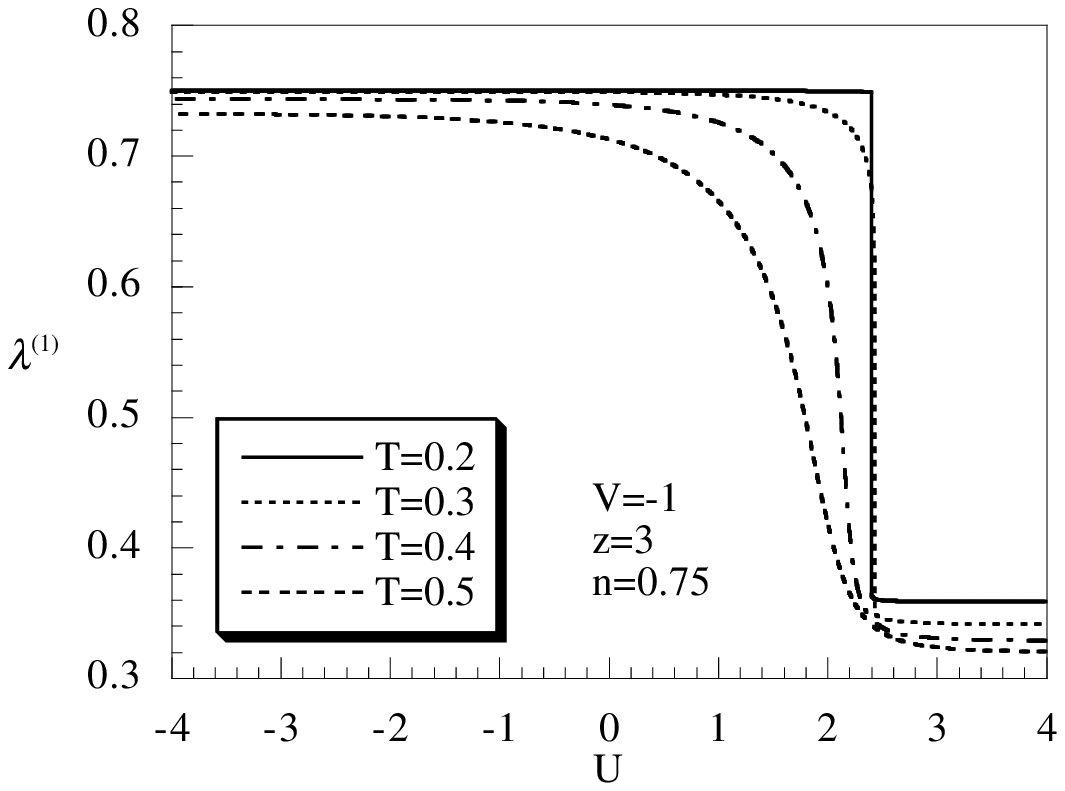}
\includegraphics[scale=0.45]{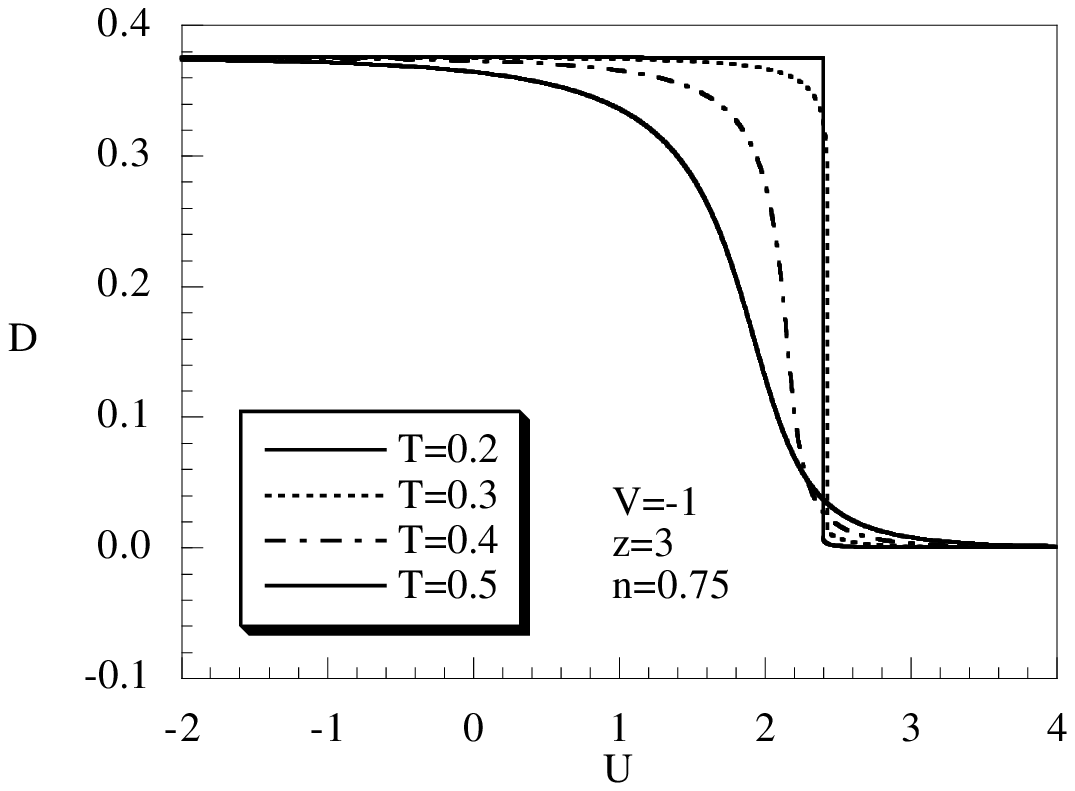}
\includegraphics[scale=0.45]{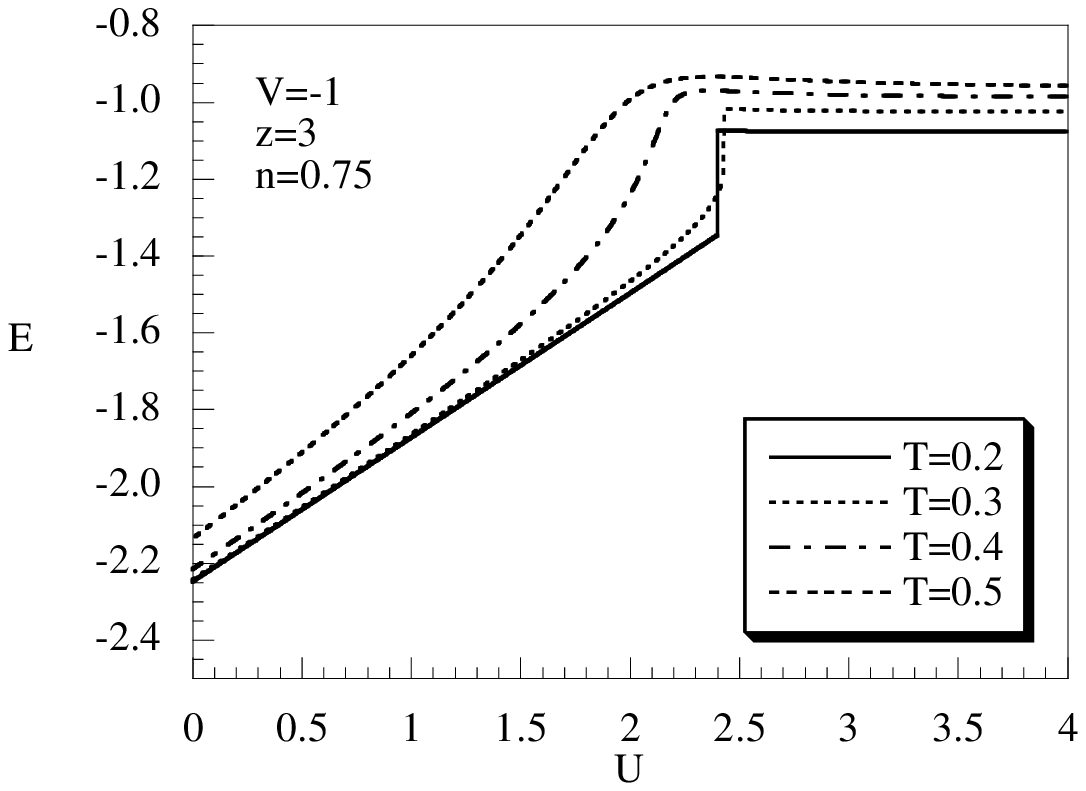}
\caption{\label{fig6}(a) The short-range correlation function $\lambda ^{\left( 1\right)}$, (b) the
 double occupancy $D$ and
(c) the internal energy $E$ as functions of $U$ for $V=-1$, $z=3$,
$n=0.75$ and for $T=0.2$, 0.3, 0.4, 0.5. }
 \end{figure}

\begin{figure}[t]
 \centering
 \subfigure[]
   {\includegraphics[scale=0.5]{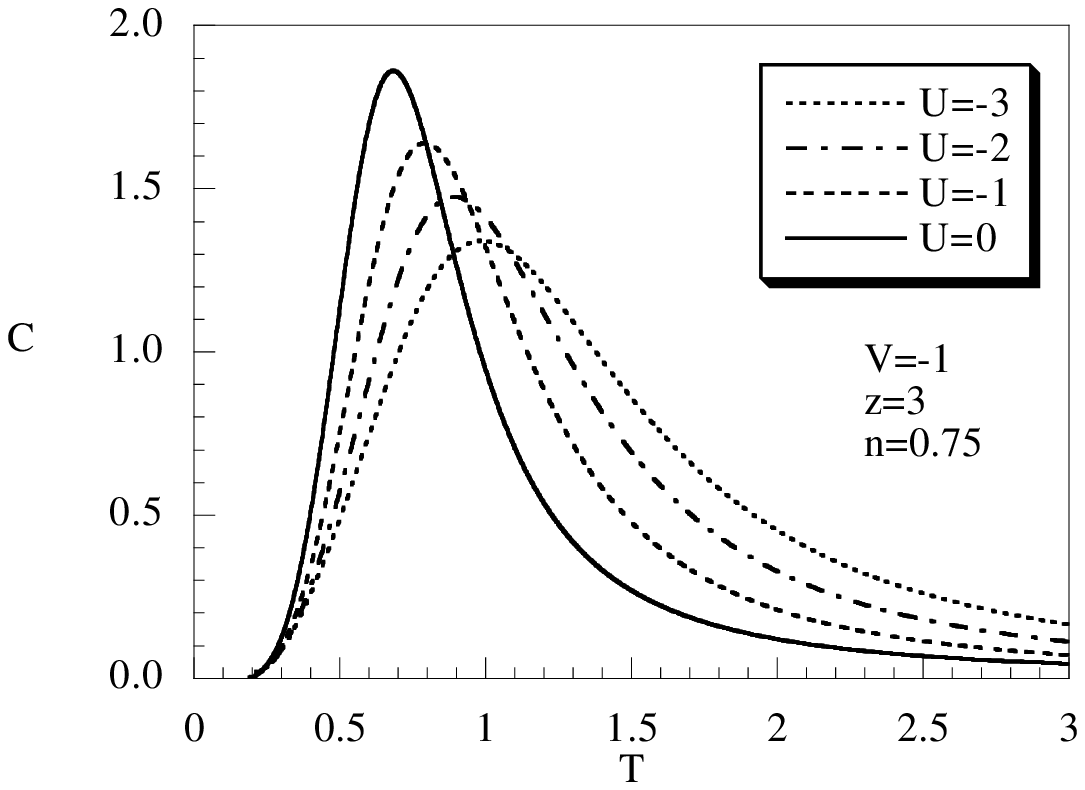}}
 \subfigure[]
   {\includegraphics[scale=0.5]{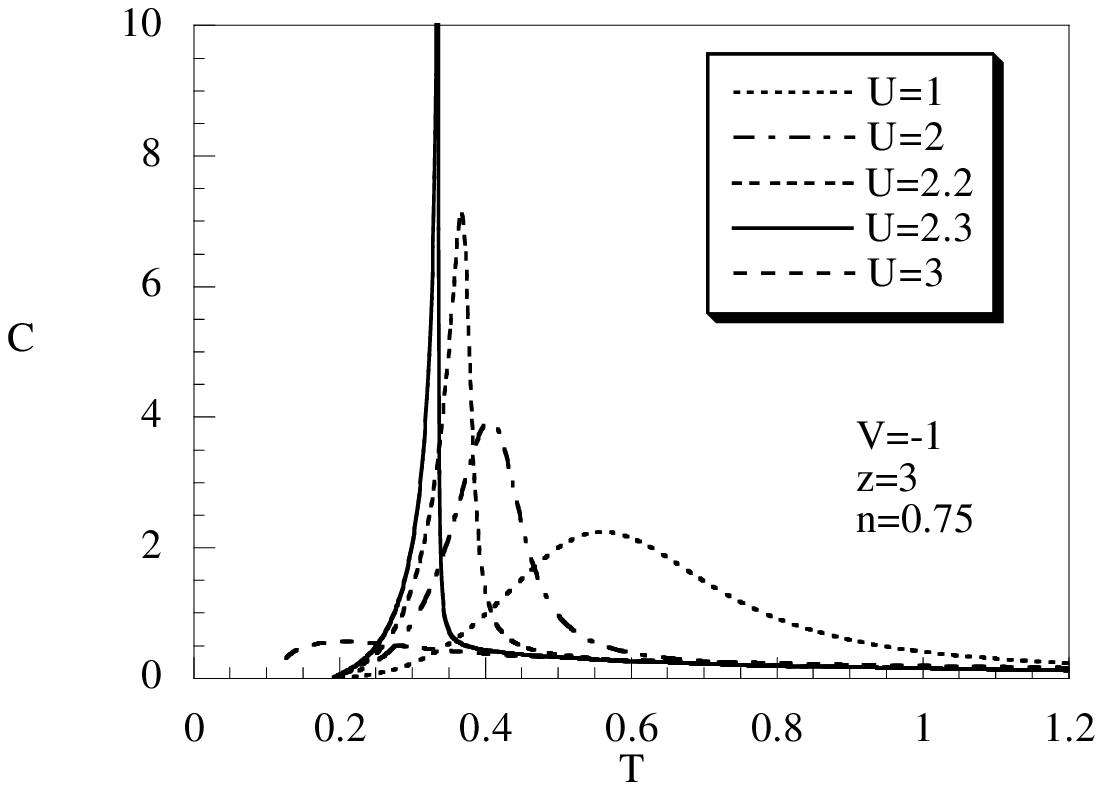}}
\caption{\label{fig7} The specific heat $C$ as a function of the
temperature $T$ for $V=-1 $, $z=3$, $n=0.75$ and for (a) $U=-3$,
-2, -1, 0; (b) $U=1$, 2, 2.2, 2.3, 3.}
\end{figure}

As it has been already pointed out, below the critical temperature
$T_{c}$ the system is thermodynamically unstable. However,  the
behavior of relevant thermodynamic quantities below $T_c$ unveils
the existence of a critical value of the on-site potential
separating the two types of phase separated configurations
sketched in Fig. \ref{fig1} \cite{mancini08_2}. Interestingly,
this critical $U$ has the same value of $U_{PS}$, turning point in
the $T_c$-curve at half filling (see Fig. \ref{figTc_vs_z}). In Figs.
\ref{fig6}, we plot the double occupancy $D$, the short-range
correlation function $\lambda ^{(1)}$ and the internal energy $E$
as functions of $U$ for $n=0.75$ and different values of the
temperature.
One observes that $D$ and $\lambda^{(1)}$ exhibit two plateaus at
low temperatures. For $z=3$, in the limit $T \to 0$, there is a
discontinuity around $U_{PS}\approx 2.3 \vert V \vert$. For
$U>U_{PS}$, the double occupancy vanishes, whereas $\lambda ^{(1)}$
tends to $n/2$. The repulsion between the electrons on the same
site and the concomitant nearest-neighbor attraction, leads to a
scenario where the electrons tend to cluster together occupying
neighboring sites. At half filling all sites are singly occupied,
whereas for $n<1$ one observes two separated clusters of filled
and empty sites. When $U<U_{PS}$, one observes a dramatic increase
(step-like as $T \to 0$) of the double occupancy and of the
short-range correlation function, namely: $D \to n/2$ and $\lambda
^{(1)}\to n$. As a consequence, the sites are doubly occupied, and
$\lambda ^{(1)}=n$ indicates that the doublons (the charge
carriers of doubly occupied sites) tend to occupy nearest-neighbor
sites arranging to form large domains occupied, leaving the rest
of the lattice empty. For $U<U_{PS}$ one also observes  a dramatic
decrease of the internal energy $E$, with a discontinuity as $T
\to T_c $ around $U\approx U_{PS}$.
In Figs. \ref{fig7}, we plot the specific heat $C$ as a function of
the temperature $T$ at $n=0.75$ and for several values of the
on-site potential $U$. For attractive $U$, the specific heat
presents one peak whose height decreases by decreasing $U$ (see
Fig. \ref{fig7}a). In Fig. \ref{fig7}b, one observes that for
repulsive $U$ the height of the peak increases by increasing $U$
moving to lower temperatures. In the limit $U \to U_{PS}$, the peak
becomes sharper and a divergence is observed at $U=U_{PS}$,
confirming the emergence of a critical point.
The same analysis carried out for different values of the particle
density, evidences a similar behavior of $D$, $\lambda ^{(1)}$ and
$C$ as functions of $U$.

\section{Repulsive intersite potential}
\label{sec_IV}

A repulsive intersite interaction disfavors the occupation of
neighboring sites. At low temperatures, this may lead to a CO
phase characterized by a distribution of the electrons in
alternating shells. In order to capture this phase, we shall solve
the self consistent equations \eqref{eq_nD_1} and \eqref{eq_nD_2}
by releasing the translational invariance requirement. To this
end, one can divide the lattice into two sublattices: $A$ contains
the central point ($0$) and the even shells, the sublattice $B$
contains the odd shells. Then, one requires the following boundary
condition (BC) to hold:
\begin{subequations}
\begin{equation}
\label{eq1} \langle n(i) \rangle =\left\{ {{\begin{array}{*{20}c}
 {n_A } \hfill \\
 {n_B } \hfill \\
\end{array} }} \right.\quad {\begin{array}{*{20}c}
 {i\in A} ,\\
 {i\in B}, \\
\end{array} }
\end{equation}
\begin{equation}
n=\frac{1}{N} \sum_i \langle n(i)\rangle= \frac{1}{2}(n_{A}+n_B).
\end{equation}
\label{boundcond}
\end{subequations}
Let us take two distinct sites $i\in A$ and $j\in B$. We require
that the expectation values of the particle density and of the
double occupancy operators at the site $i$ are equal to the ones
of the neighboring sites of $j$ and viceversa, namely:
\begin{equation}
\label{eq11}
\begin{split}
  \langle n(i)\rangle&= \langle n^\alpha (j)\rangle ,\quad \quad
  \langle n(j)\rangle= \langle n^\alpha (i)\rangle,\\
  \langle D(i)\rangle&= \langle D^\alpha (j)\rangle,\quad \quad
 \langle D(j)\rangle =  \langle D^\alpha (i)\rangle
 .
  \end{split}
\end{equation}
Thus, by means of Eqs. \eqref{eq_nD_1} and \eqref{eq_nD_2}, one
finds two equations
\begin{equation}
\label{eq12}
\begin{split}
&\frac{2 ( f F_A^z+gG_A^z)}{1+2f F_A^z+g
G_A^z}=
\frac{1}{1+2f F_B^z+g  G_B^z}
\left[ X_B     \right.
\\
&+ \left.  2fK(X_B+2aY_B)F_B^{z-1}+ g
K^2(X_B+2dY_B)G_B^{z-1} \right],
\\
&\frac{g  G_A^z}{1+2f F_A^z+g  G_A^z} =
\frac{Y_B(1+2fK^2F_B^{z-1}+g  K^4 G_B^{z-1})}{1+2f F_B^z+g G_B^z},
\end{split}
\end{equation}
plus two more equations which can be obtained by substituting $A
\leftrightarrow B$. The coefficients subscripts pertain to sites
belonging to the two different sublattices. In order to close the
set of self-consistent equations one needs one more equation. By
using the BC \eqref{boundcond}, one can fix the particle density
as
\begin{equation}
\label{eq13} n=\frac{f F_A^z+gG_A^z}{1+2f F_A^z+g
G_A^z}+\frac{fF_B^z+gG_B^z}{1+2f F_B^z+g G_B^z}.
\end{equation}
As a result, one can now determine all the unknown parameters $X_A$,
$X_B$, $Y_A$, $Y_B$, and $\mu$. By means of the same analysis
employed in the previous section, one can express the local
correlators $\kappa^{(p)}$ and $\lambda^{(p)}$ in terms of these
parameters, and eventually compute all the properties of the
system. In particular, one finds:
\begin{equation}
\label{eq915}
\begin{split}
D &=\frac{1}{2}\left( D_A +D_B \right)  \\
&=\frac{g  G_A ^z}{2(1+2f F_B ^z+g G_A^z)}+
\frac{g  G_B ^z}{2(1+2f F_B ^z+g  G_B ^z)},\\
 \lambda ^{(1)}&=\frac{ f K(X_A +2aY_A )F_A ^{z-1}+g
 K^2(X_A +2dY_A )G_A ^{z-1}}{1+2f F_{B} ^z+g  G_B ^z}.
\end{split}
\end{equation}
By varying the external parameters $U$, $n$ and $T$ one has the
tools to completely characterize the phase diagram pertinent to
the case of repulsive intersite interactions. In the following we
shall set $V=1$.

\begin{figure}[t]
\centering
 \subfigure[]
   {\includegraphics[scale=0.6]{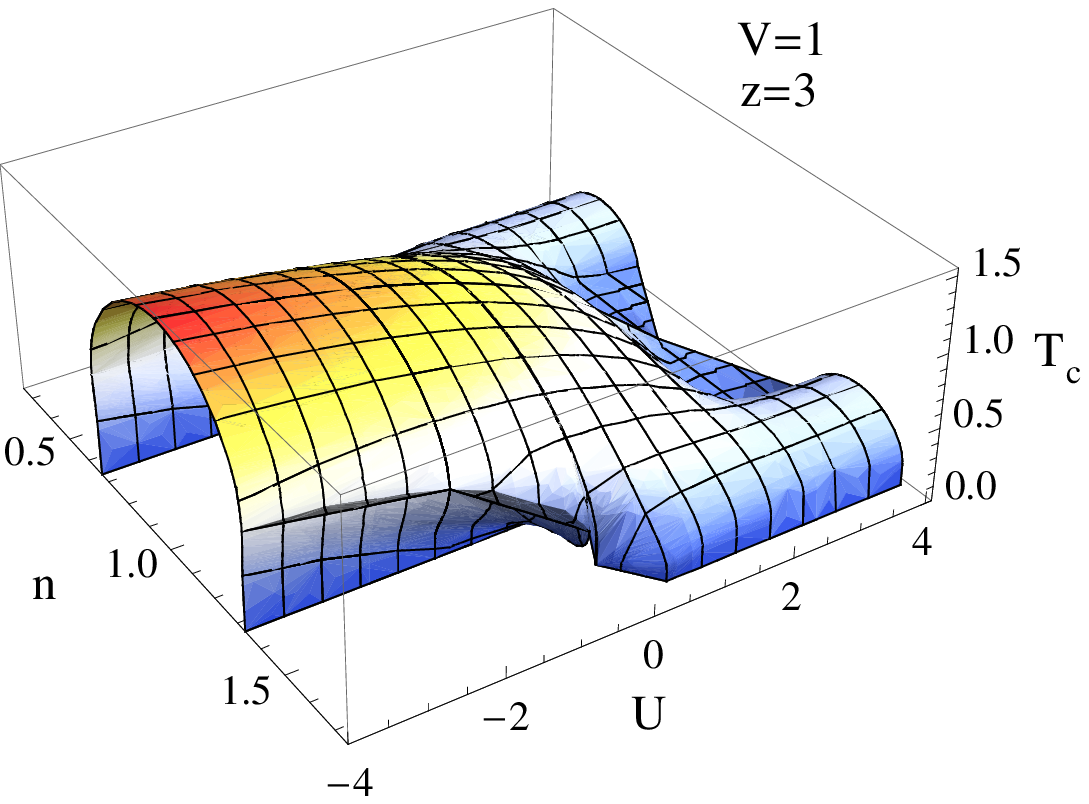}}
    \subfigure[]
   {\includegraphics[scale=0.225]{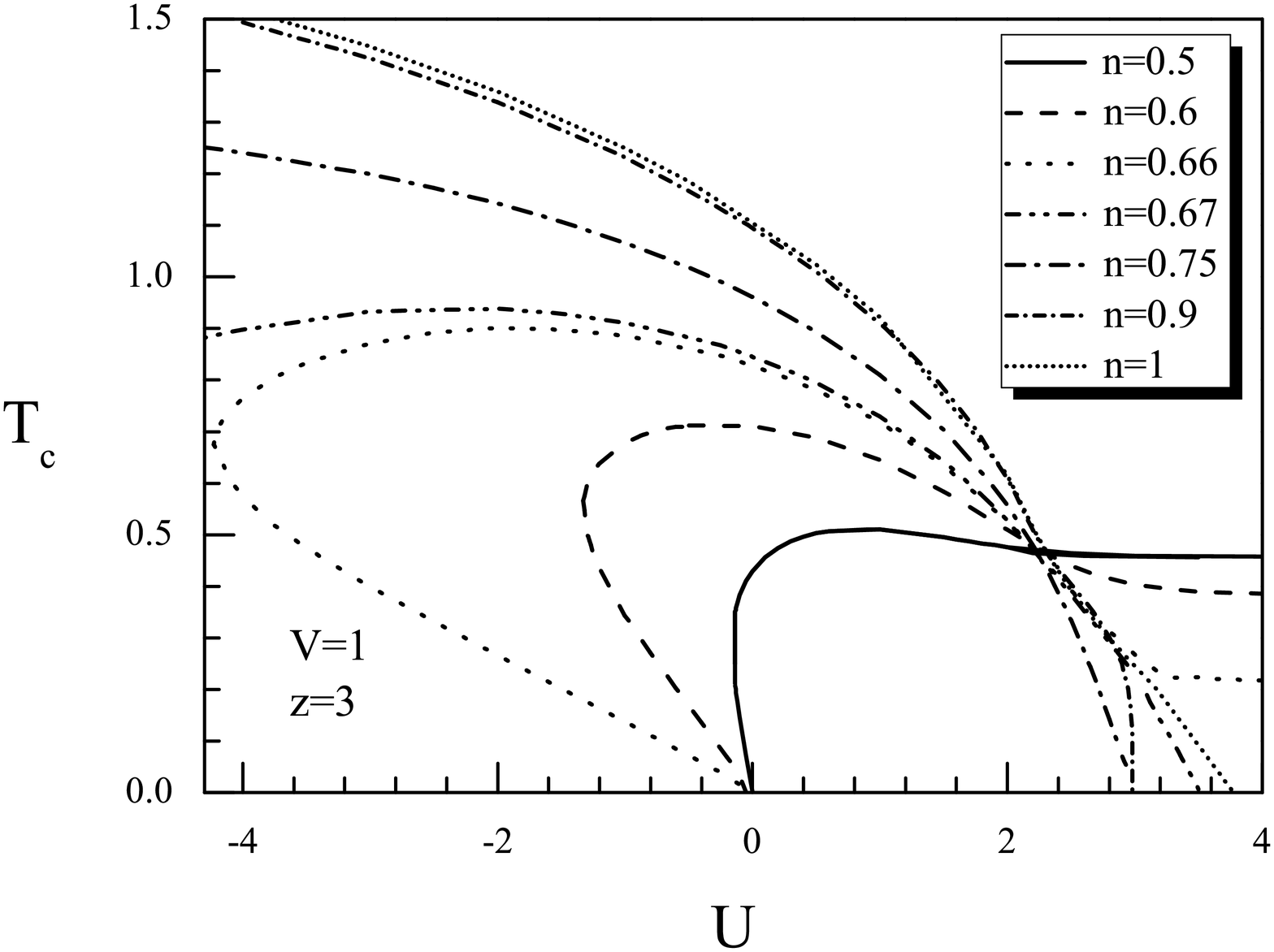}}
\caption{\label{phadiamV1}  (Color online) (a) Phase diagram in the
space ($U,n,T$) for $V=1$ and $z=3$. (b) Phase diagram in the
plane ($U,T$) for $V=1$ and $z=3$  and several values of $n$.}
\end{figure}

\subsection{Phase diagram}

\begin{figure}[t]
 \centering
\includegraphics[scale=0.56]{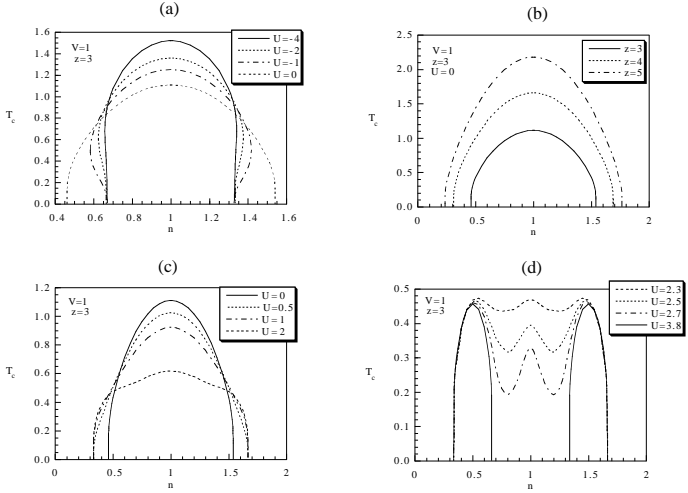}
\caption{\label{Tc_vs_n} Phase diagram in the plane ($T,n$) for
$V=1$: Figs. \ref{Tc_vs_n}a, \ref{Tc_vs_n}c, \ref{Tc_vs_n}d  $z=3$
and several values of $U$; Fig. \ref{Tc_vs_n}b, $U=0$ and
$z=3,4,5$.}
 \end{figure}

In this section, we derive the phase diagram: by numerically
solving the set of equations \eqref{eq12} and \eqref{eq13} we find
a region of the $(U,n,T)$ 3D space characterized by a spontaneous
breakdown of the translational invariance. In this region, shown
in Fig. \ref{phadiamV1}a for the case $z=3$, the population of the
two sublattices $A$ and $B$ is not equivalent: the system has
entered a finite temperature long-range CO phase. Upon decreasing
the temperature, the distribution of the electrons becomes more
and more inhomogeneous. To better understand the typology of the
critical region we take sections of the 3D structure and study the
2D phase diagrams at constant $n$ in Fig. \ref{phadiamV1}b and at
constant $U$ in Fig. \ref{Tc_vs_n}.

In the plane $(U, T)$, the  critical temperature line shows
different behaviors according to the value of the particle
density. As illustrated in Fig. \ref{phadiamV1}b for the case
$z=3$, one encounters the following situation. (i) For $n<1/z$
there is no CO phase for all values of $U$; (ii) For $1/z<n<n_p$ a
CO phase is observed only for $U>0$ ($n_p=0.46$ for $z=3$); the
critical temperature $T_c$ increases by increasing $U$ and tends
to a constant (depending on the value of $n$) in the limit  $U \to
\infty$. (iii) For $n_p<n<2/z$ there is a CO phase for both
attractive and repulsive $U$, with a reentrant behavior for $U<0$.
(iv) For $n>2/z$ there is no reentrant behavior: from a constant
value at large negative $U$, the critical temperature decreases by
increasing $U$ and vanishes at a certain value of $U$. An
interesting feature is the presence of a crossing point in the
critical temperature curves around $U=2.3V$ for $z=3$.
In the plane $(n,T)$, at fixed $U$, the CO phase is observed in
the interval $n_1<n<n_2$; the width $\Delta=n_2-n_1$ varies with
the temperature, following different laws according to the value
of $U$. At $T=0$ a complete CO state is established in the regions
$2/z \le n \le 2(z-1)/z$  for $U<0$ and $1/z \le n \le (2z-1)/z$
for $U>0$. For $U<0$ (see Fig. \ref{Tc_vs_n}a), $\Delta n$ first
increases with $T$, then decreases vanishing at $n=1$, where the
maximum critical temperature is reached; a reentrant behavior
characterizes this region. As it is shown in Fig. \ref{Tc_vs_n}b,
the value $U=0$ is a singular point; at $T=0$ the CO phase exists
in the interval $n_p<n<2-n_p$. For  $0<U \le 2$ (see Fig.
\ref{Tc_vs_n}c), $\Delta n$ decreases with $T$ and no reentrant
phase is observed; as in the attractive case, the phase diagram
presents a single lobe structure centered at $n=1$. For $U>2$ (see
Fig. \ref{Tc_vs_n}d), one observes the formation of two lobes
centered around $n=0.5$ and $n=1.5$ and the corresponding
decreasing of the central lobe centered at $n=1$, which disappears
at $U=U_0$ ($U_0 \approx 3.7$  for $z=3$). For $U>U_0$ the two
lobes are separated; at $T=0$ the CO phase is observed in two
regions centered at $n=0.5$ and $n=1.5$. It is worthwhile to
notice that for $U<2V$ and for all values of $T$ , $\Delta n$
increases by increasing the coordination number $z$; in the limit
$z \to \infty$, $\Delta n \to 2$  and the CO phase is observed for
all values of $n$. Besides the translational-invariance broken
solution in the 3D region illustrated in Fig. \ref{phadiamV1}a,
the set of equations \eqref{eq12} and \eqref{eq13} admits also a
homogeneous solution. In order to determine which solution is
energetically favored one has to look at the free energy. The
Helmholtz free energy $F$ can be computed by means of the formula
 \begin{figure}[t]
 \vspace{-8mm}
 \centering
 \subfigure[]
   {\includegraphics[scale=0.5]{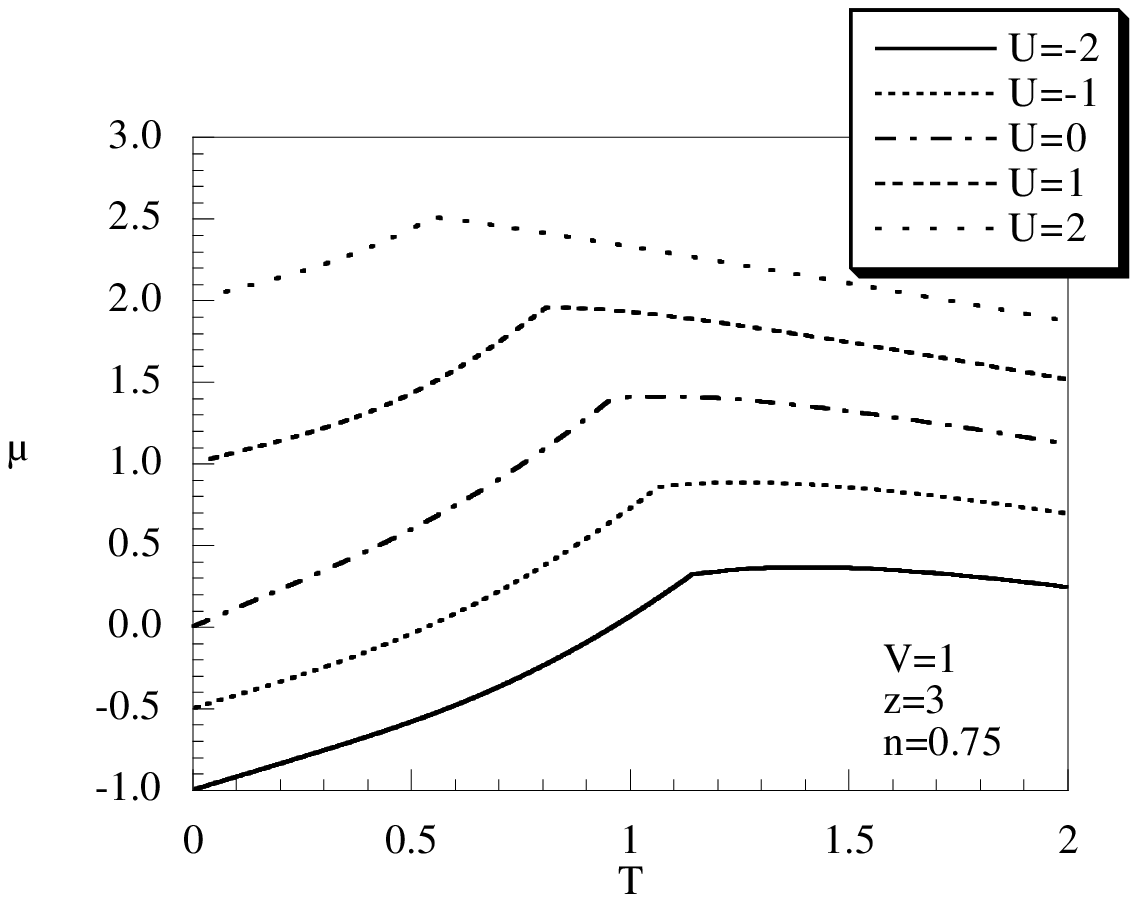}}
 \subfigure[]
   {\includegraphics[scale=0.5]{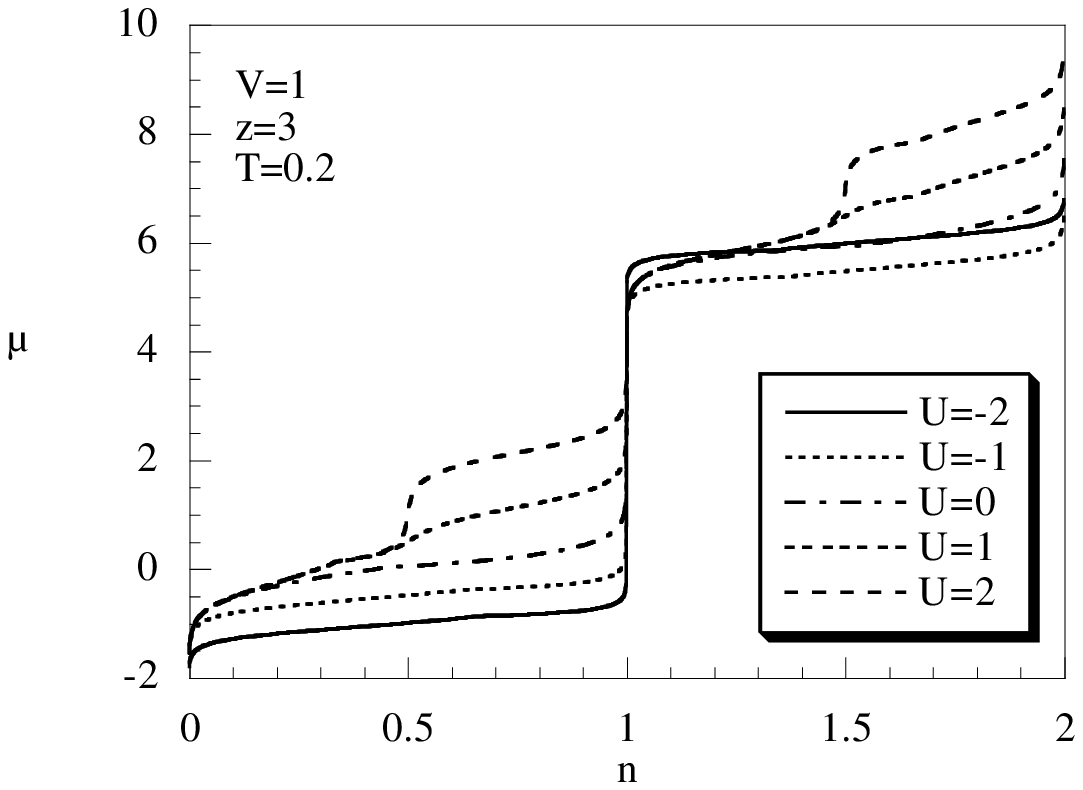}}
\caption{\label{fig11} (a) The chemical potential $\mu$ as a
function of the temperature for $V=1$, $z=3$, $n=0.75$ and different
on-site interactions. (b) The chemical potential $\mu$ as a
function  of the particle density $n$ for $z=3$, $T=0.2$ and
various on-site interactions.}
 \end{figure}
\begin{equation*}
F(T,n)=\int\limits_0^n \mu (T,n')dn' .
\end{equation*}
Upon defining the difference $\Delta F=F_{hom} -F_{CO}$ - where
$F_{hom} $ is the free energy of the homogeneous phase and
$F_{CO}$ the one of the CO phase, respectively -  one finds that below the
transition temperature
the CO phase is energetically favored since $\Delta F\ge 0$. In
particular, $\Delta F$ smoothly vanishes for $T\to T_c $,
signalling a second-order phase transition.

\subsection{Thermodynamic properties}

In this section, we shall determine several thermodynamic
quantities whose behaviors support the scenario depicted in the
previous section. The behavior of the chemical potential as a
function of the temperature and of the particle density $n$ is
reported in Figs. \ref{fig11}a and \ref{fig11}b, respectively. One
can immediately notice that, as a function of the temperature,
$\mu$ presents a cuspid at $T=T_c$: for $T \ge T_c$ ($T \le T_c$)
$\mu$ is a decreasing (increasing) function of $T$. When plotted
at fixed temperature, the chemical potential is always an
increasing function of $n$, hinting at a thermodynamically stable
system, for both the CO and homogeneous phases. In the limit  $T
\to 0$, $\mu$ shows two plateaus for $U<2V$ in the range $0<n\le
2$; when $U>2V$ each plateau splits in two sub-plateaus.

\begin{figure}[t]
 \centering
 \subfigure[]
   {\includegraphics[scale=0.5]{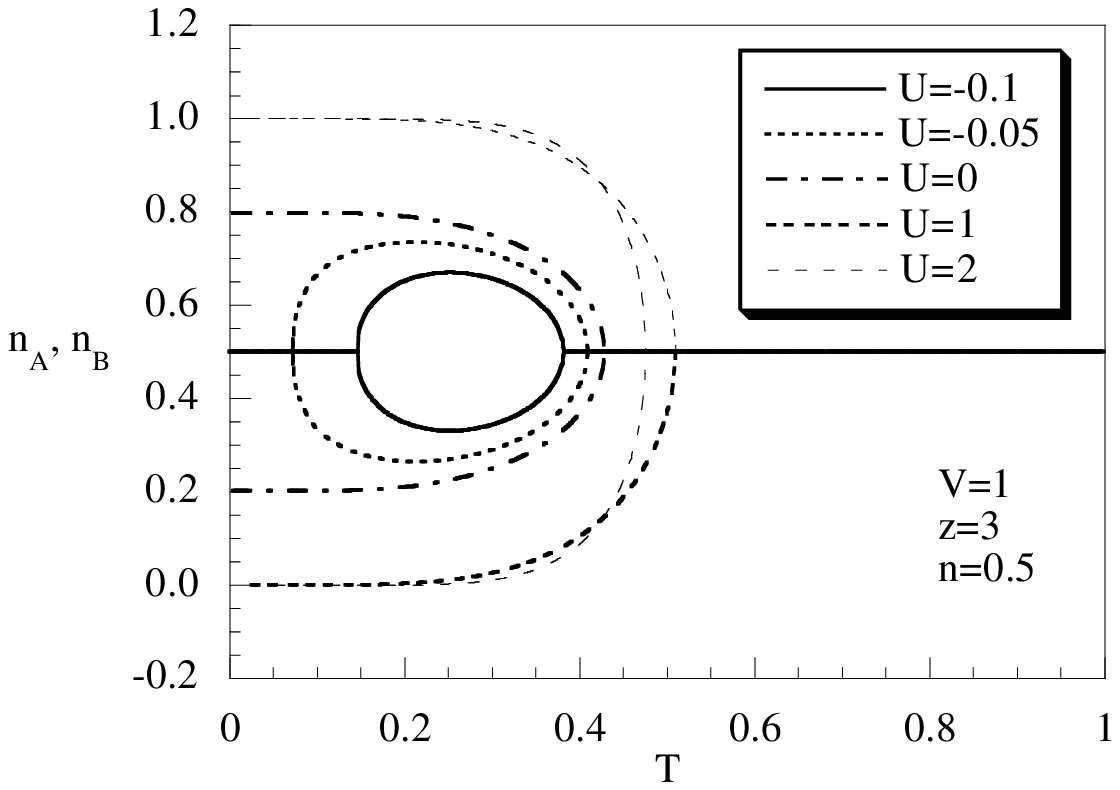}}
 \subfigure[]
   {\includegraphics[scale=0.5]{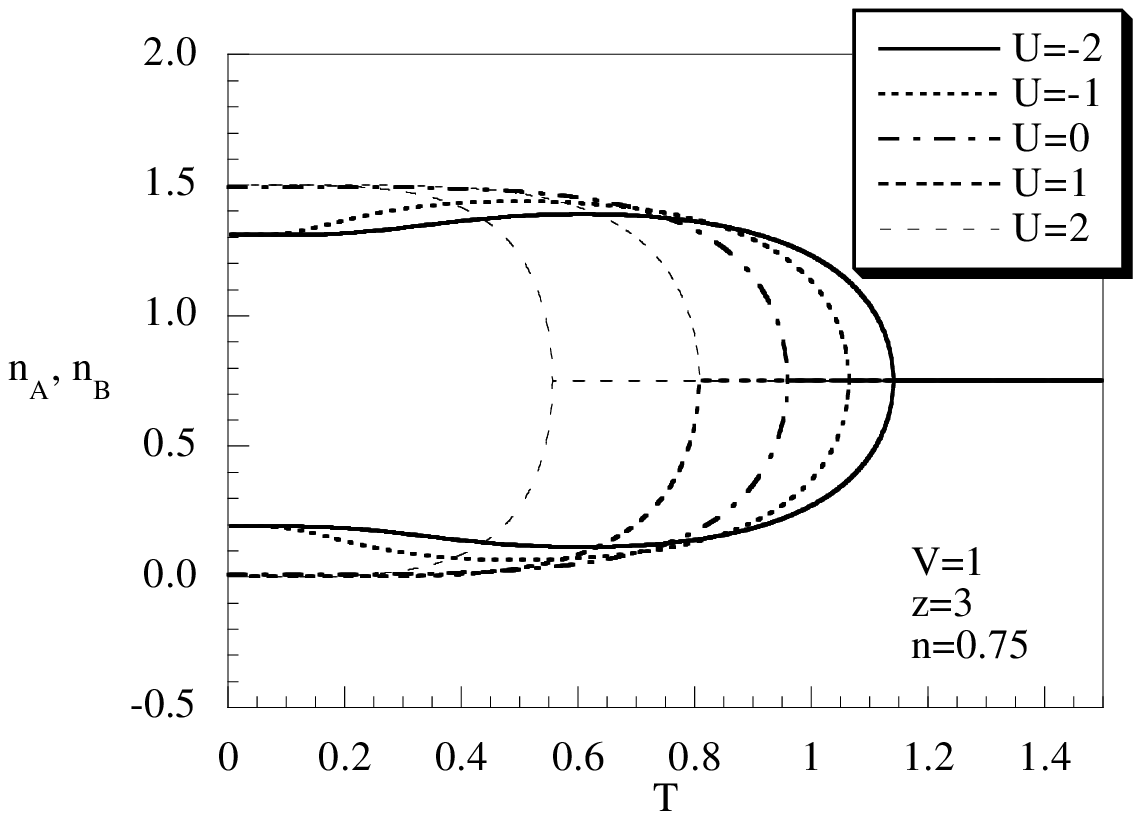}}
 \caption{\label{fig12} The sublattices variables $n_{A}$
and $n_{B}$ as functions of the temperature for $V=1$, $z=3$,
several values of $U$ and: (a) $n=0.5$,  (b) $n=0.75$.}
 \end{figure}

A full comprehension of the phase diagram and of the distribution
of the particles on the sites of the Bethe lattice can be achieved
by a detailed investigation of the particle density and of the
double occupancy. In particular, the study of  the different
contributions coming from the two sublattices, reveals the onset
of CO states, the reentrant behavior, and also the distribution of
the electrons in the shells of the Bethe lattice. In Figs.
\ref{fig12}a and \ref{fig12}b, we plot the sublattices variables
$n_{A}$ and $n_B$ as functions of the temperature  for $n=0.5$,
$n=0.75$, and several values of $U$.
 At high temperatures the
particles distribute homogeneously in the entire lattice:
$n_{A}=n_B=n$. Upon lowering the temperature, one finds that at
the critical temperature $T_c=T_c(n,U)$ a CO phase is established;
the particles tend to fill one sublattice ($B$) and to empty the
other sublattice ($A$). By further lowering the temperature, the
behavior is different according to the values of $n$ and $U$.
For $n=0.5$, one observes the following situation: (i) for $U$
negative and close to zero the phase diagram exhibits a reentrant
behavior (see Fig. \ref{phadiamV1}b); correspondingly, as shown in
Fig. \ref{fig12}a, there is another critical temperature at which
the translational invariance is restored and the two sublattices
are again equally populated; (ii) for $U=0$, the anisotropy of the
filling increases and, at $T=0$, one finds $n_A=2n/5$  and
$n_B=8n/5$ for $z=3$; (iii) for $U>0$ a total charge order is
established in the limit $T \to 0$: all the particles reside in one
sublattice ($B$) whilst the other is completely empty. For
$n=0.75$ and $n=1$ (not shown), there is no reentrant phase; by
lowering $T$,  $n_A$ ($n_B$) decreases (increases) and tends to
zero (2$n$) at $T=0$ (for $n=0.75$ and $U<0$ charge ordering is not
complete, since a small fraction of particles is found in the
sublattice $A$, also at $T=0$).

\begin{figure}[t]
 \centering
 \subfigure[]
   {\includegraphics[scale=0.5]{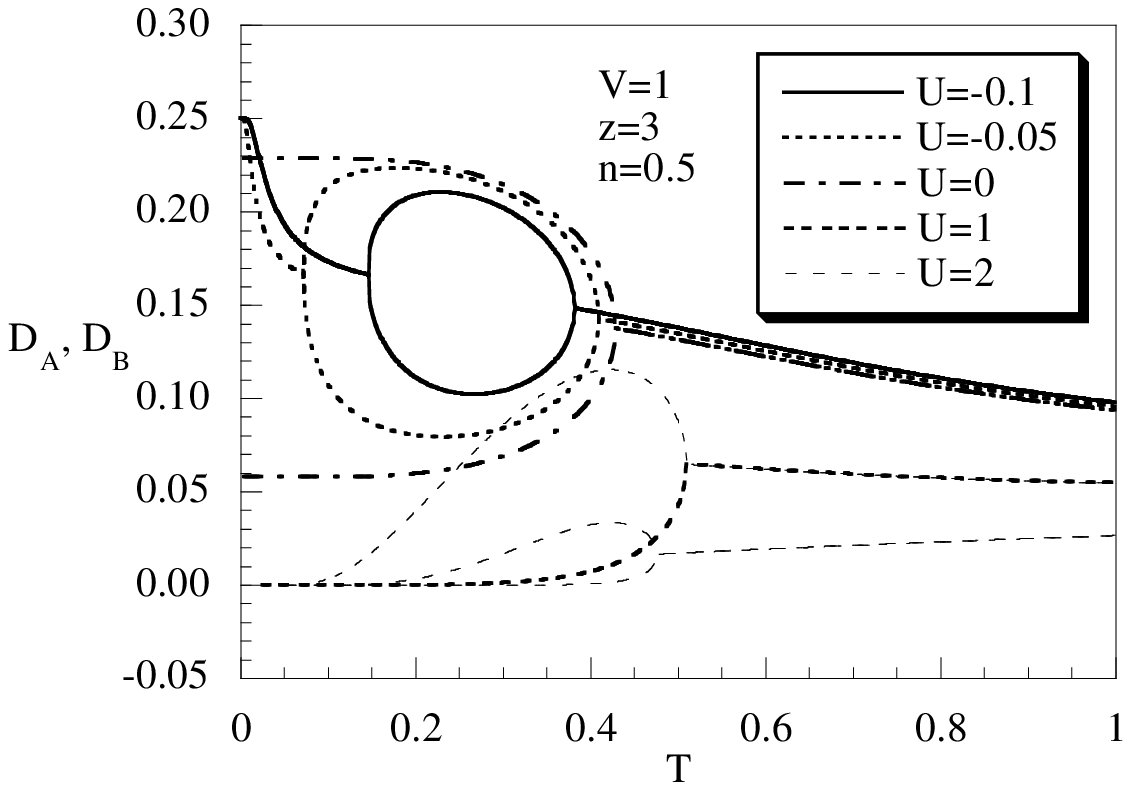}}
 \subfigure[]
   {\includegraphics[scale=0.5]{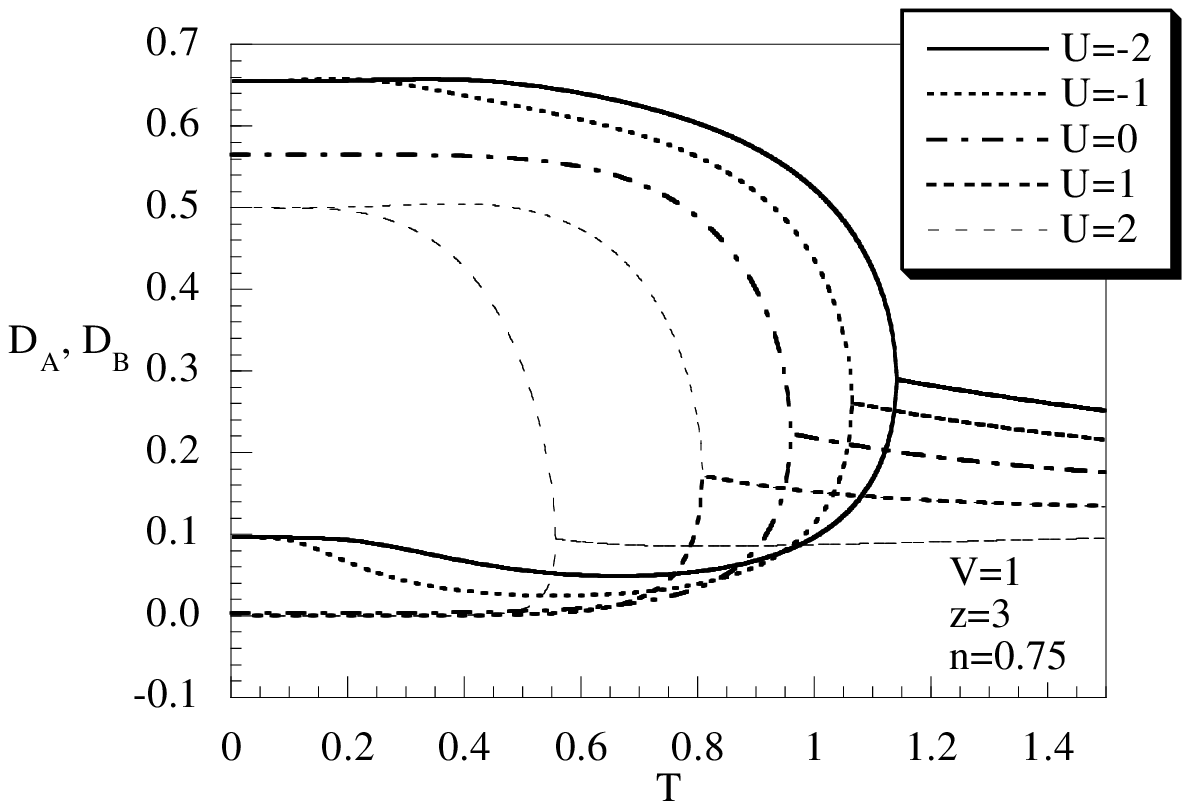}}
\caption{\label{fig13} The sublattices variables $D_{A}$ and
$D_{B}$ as functions of the temperature as functions of the
temperature for $V=1$, $z=3$, several values of $U$ and: (a)
$n=0.5$, (b) $n=0.75$.}
 \end{figure}

In Figs. \ref{fig13}a and  \ref{fig13}b, we plot the sublattice
double occupancies $D_{A}$ and $D_{B}$ as functions of the
temperature for $n=0.5$ and $n=0.75$ and several values of $U$. At
high temperatures the system is in a homogeneous phase and,
correspondingly $D_A=D_B=D$. Upon lowering the temperature, one
finds that at $T_c$ there is a phase transition to a CO state:
$D_B$  increases while $D_{A}$  decreases. By further lowering the
temperature, one observes different behaviors according to the
values of $n$ and $U$. A reentrant phase characterized by a second
critical temperature at which the system returns to the
homogeneous phase where the double occupancy tends to $n/2$ in the
two sublattices (observed for $n=0.5$ and $U$ negative and small).
The CO persists in the limit $T \to 0$ with: (i)  $D_{A}$ and
$D_{B}$ both vanishing at $T=0$ (observed for $n=0.5$  and $U>0$);
(ii) $D_{A}$ ($D_B$) decreases (increases) and tends to a finite
value (observed for $n=0.75$  and $U<0$); (iii) $D_{A}$ ($D_B$)
decreases (increases) and tends to zero ($n$) (observed for
$n=0.75$, $U>0$ and $n=1$, $\forall U$).


Since the specific heat exhibits a very rich structure in
correspondence to the critical lines in the phase diagram shown in
Fig. \ref{phadiamV1}a, we shall investigate its behavior for
different values of the particle density $n$. The possible
excitations of the ground state are creation and annihilation of
singly occupied or doubly occupied states, induced by the Hubbard
operators $\psi^{(\xi)}$ and $\psi^{(\eta)}$, respectively
\cite{mancini08_PRE}. The corresponding  transition energies are
given in Eqs. \eqref{energies} and the high and low temperature
peaks exhibited by the specific heat are due to these transitions.
One may distinguish between them by looking at the position of the
peaks, i.e., if the position changes or remains constant by
varying $U$. Besides these peaks, one also observes peaks in
correspondence to the transition temperatures separating
homogeneous and CO phases. Of course, in the case of a reentrant
phase, one correspondingly observes two peaks relative to the
transitions.
 \begin{figure}[t]
 \centering
\includegraphics[scale=0.55]{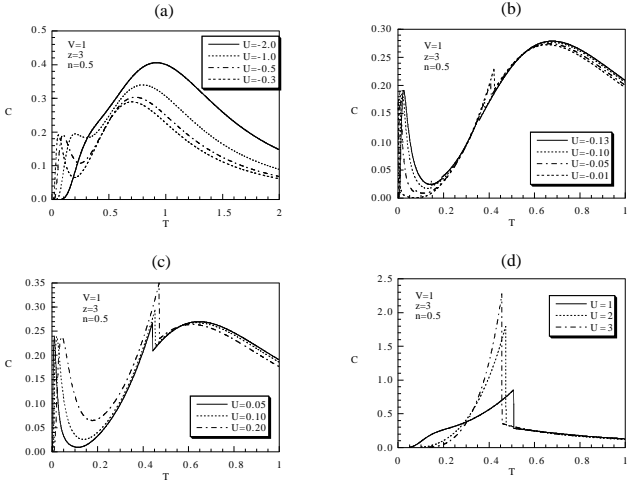}
\caption{\label{fig19}(a) The specific heat as a function of the
temperature for $z=3$, $V=1$, $n=0.5$ and different regions of
$U$: (a) attractive interaction $U=-2,\ldots, -0.3$; (b) $U$
negative and close to zero; (c) $U$ positive and close to zero;
(d) $U$ large and positive.}
 \end{figure}
 The behavior of the specific heat at $n=0.5$ as a
function of the temperature is shown in Figs. \ref{fig19}a-d. For
$U$ negative and large ($U=-3V, -2V$), there is no CO and $C$
exhibits only one peak at high temperature at position $T_1
\approx 1$; by increasing $U$ ($U=-V$, $-0.5V$, $-0.3V$) a second
peak appears at low temperatures $T_2 \approx 0.1$ and the
position of the first peak decreases (see Fig. \ref{fig19}a).
Further increasing $U$, a reentrant phase transition from the
homogeneous phase to the CO state occurs; correspondingly, two new
peaks are observed around the two critical temperatures $T_3
\approx 0.4$ and $T_4 \approx 0.15$ (see Fig. \ref{fig19}b). For
attractive on-site interactions the possible excitations are
creation and annihilation of doubly occupied states, and the
charge peaks at $T_1$ and $T_2$  are mainly  induced by
$\psi^{(\eta)}$; these peaks move towards low temperatures, with
$T_2 \to 0$  as $U \to 0$. For $U$ small and positive and close to
zero, there is a phase transition but no reentrant phase, $T_3=0$.
The specific heat exhibits a three-peak structure (see Fig.
\ref{fig19}c): one peak at high temperature ($T_1 \approx 0.7$), a
second peak at low temperatures ($T_2 \approx 0.02$), and a third
peak around $T_3 \approx 0.4$. For $U$ positive and large (see
Fig. \ref{fig19}d) only one peak at $T_3 \approx 0.5$ appears due
to the phase transition to the CO state. The behavior of the
specific heat at $n=0.75$ as a function of the temperature is
shown in Figs. \ref{fig20}a-c. For $n=0.75$ there is no reentrant
behavior. At low temperatures the system is in a CO state and
exhibits a phase transition at $T_c$ to a homogeneous phase.
 \begin{figure}[t]
 \centering
 \subfigure[]
   {\includegraphics[scale=0.45]{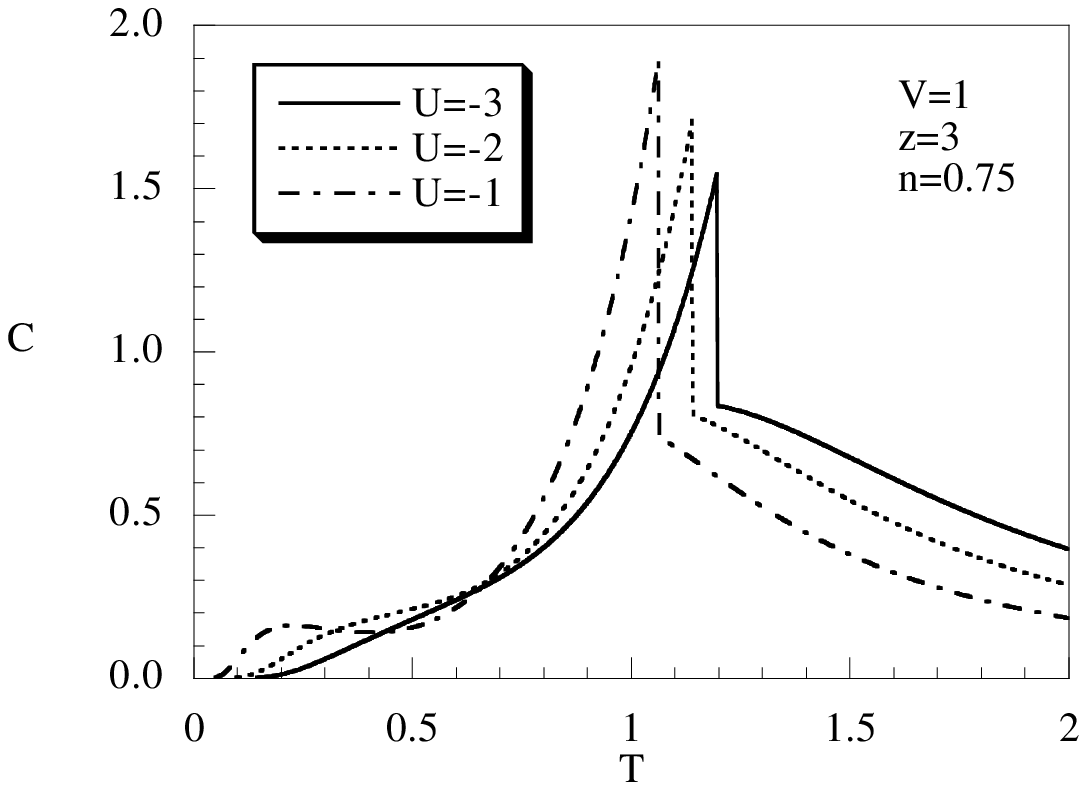}}
  \subfigure[]
   {\includegraphics[scale=0.45]{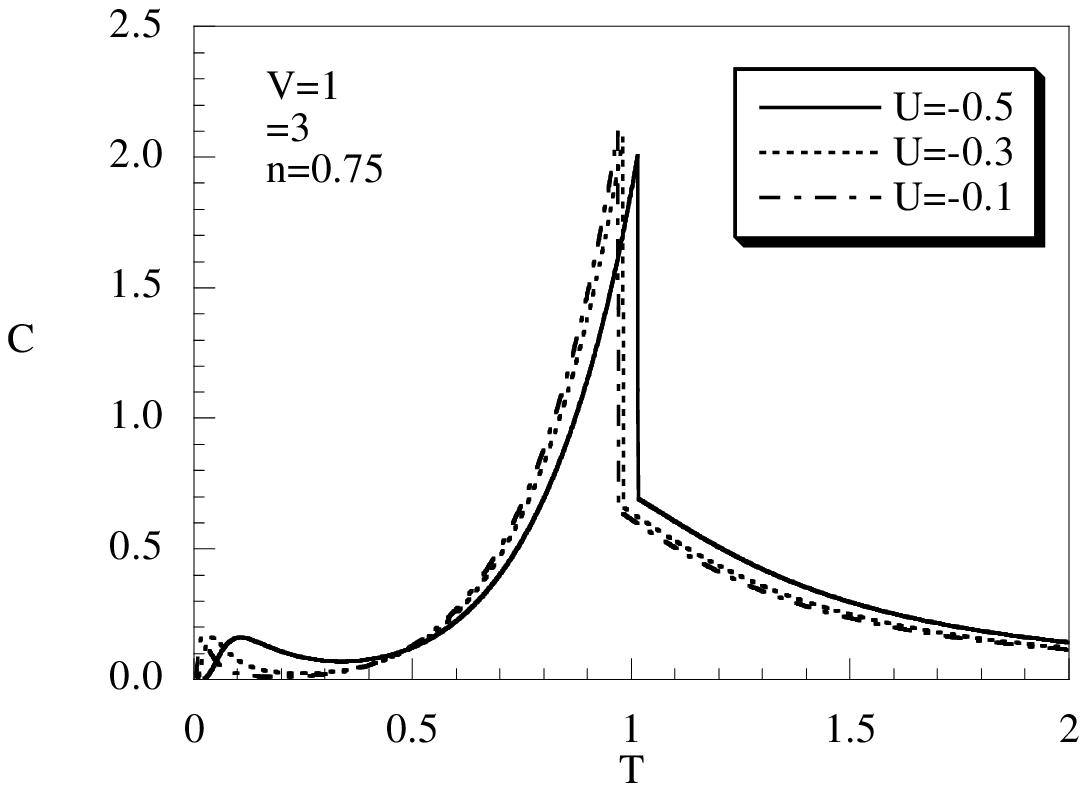}}
\subfigure[]
   {\includegraphics[scale=0.45]{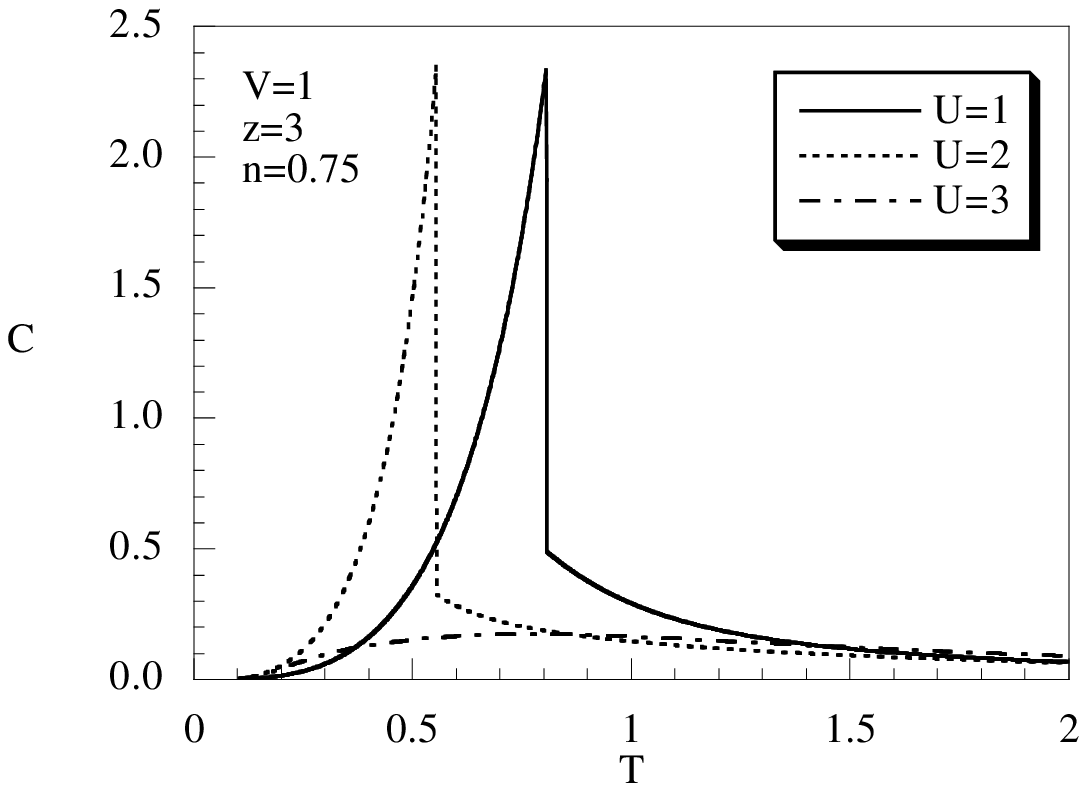}}
\caption{\label{fig20}
The specific heat as a function of the temperature for $z=3$, $V=1$, $n=0.75$
 and different regions of $U$:
(a) large attractive interaction $U=-2 , \ldots, -1$; (b) $U$ negative
and close to zero; (c) $U$ large and positive. }
 \end{figure}
For $U$ negative and large only one peak appears at $T_3=T_c$, due
to the phase transition. For $U$ negative and small, besides the
peak at  $T_3$, one observes another peak  $T_2$ at low
temperatures which vanishes as $U$ approaches zero. For  $0 \le U
<3V$, there is only one peak at $T_3=T_c$; contrarily to the case
of $n=0.5$, the peak $T_2$  is not observed for $U$ positive. For
$U \ge 3V$ ($z=3)$ there is no phase transition and one broad peak
of much more intensity is observed. The behavior of the specific
heat at half filling as a function of the temperature is shown in
Figs. \ref{fig21}a-b. For $n=1$, a CO phase is observed for all
values of  $U<U_0$. As it is shown in Fig. \ref{fig21}a, the
specific heat exhibits one peak situated at  $T_3=T_c$  moving
towards low temperatures as $U$ increases. At $T=0$, for $ U<U_0$,
one sublattice ($A$) is empty, whereas the other sublattice ($B$)
has all sites  doubly occupied; there are neither singly occupied
sites nor neighbor sites occupied ($\lambda^{(1)}=0$). The energy
of the ground state is non-degenerate. For $U>U_0$, one observes a
homogeneous state: all sites are singly occupied and the energy of
the ground state is infinitely degenerate since the spins are
arbitrarily oriented.
 \begin{figure}[t]
 \centering
 \subfigure[]
   {\includegraphics[scale=0.5]{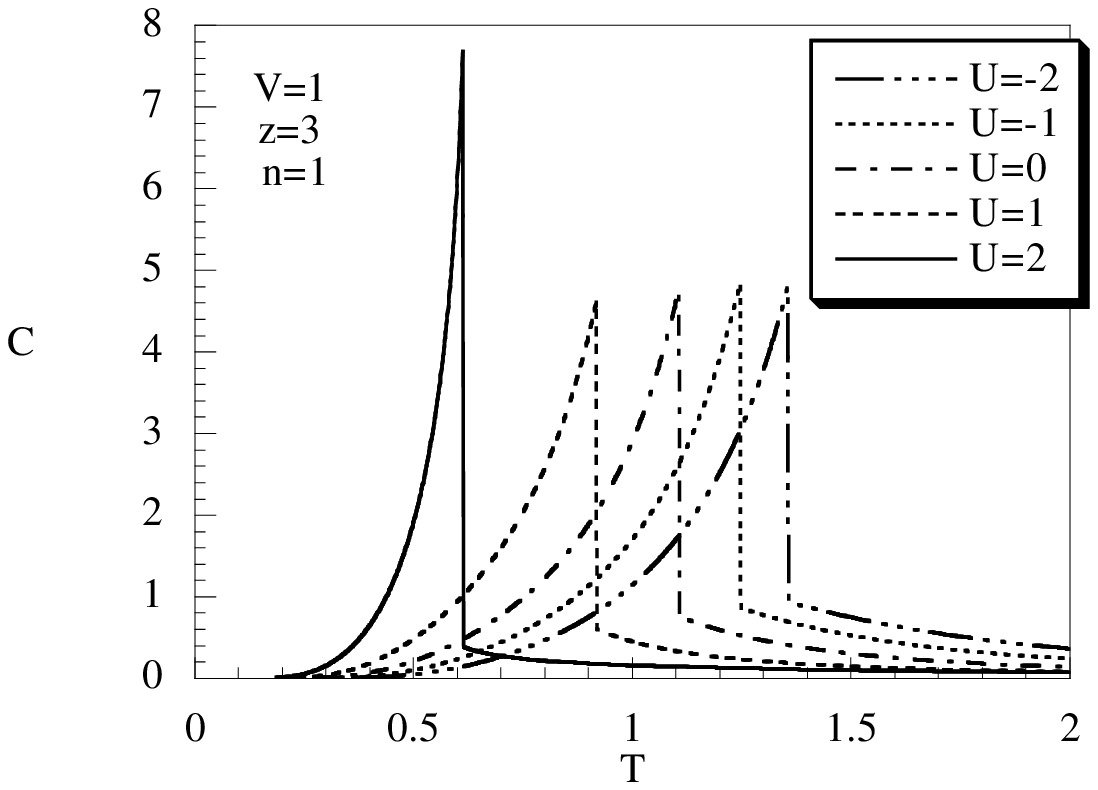}}
 \subfigure[]
   {\includegraphics[scale=0.5]{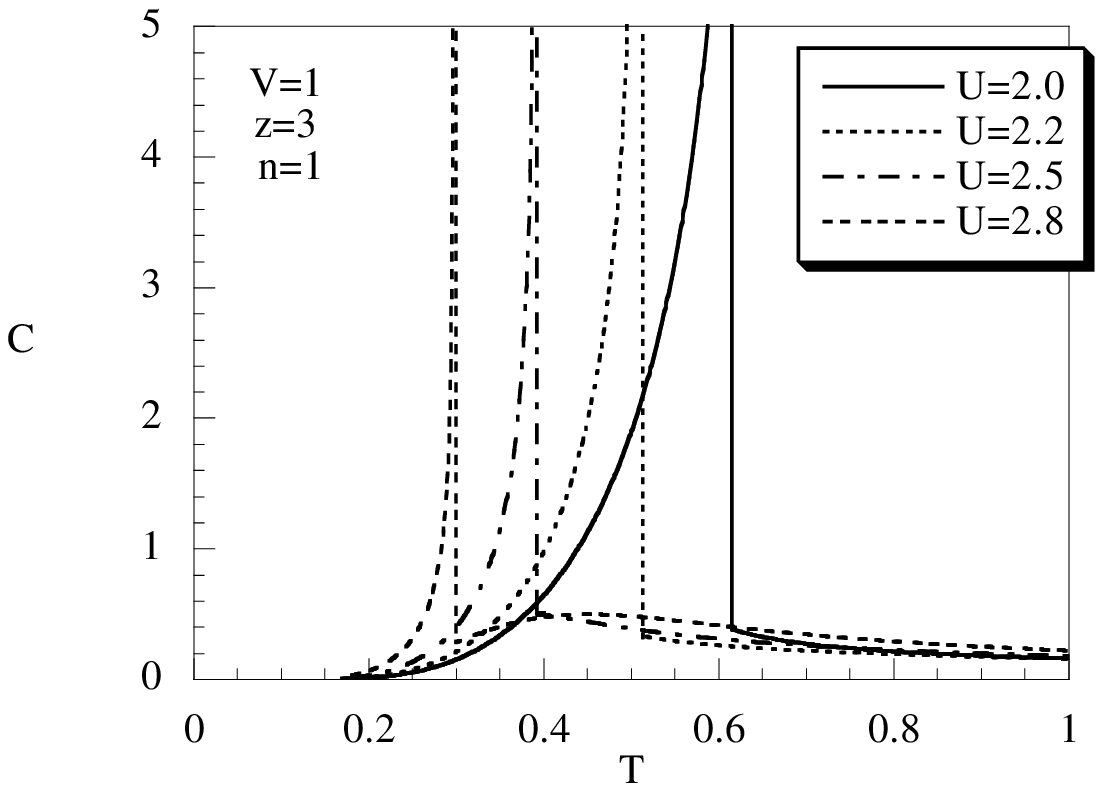}}
\caption{\label{fig21}The specific heat as a function of the
temperature for $z=3$, $V=1$, $n=1$ and different regions of $U$:
(a) values of $U$ far from the critical point  $U_0$;
(b) values of $U$ approaching $U_0$.}
 \end{figure}
At $T=0$ and $U=U_0$, there is a phase transition from a
non-degenerate level to a degenerate one; therefore one expects a
discontinuity in the internal energy  $E$ and, correspondingly, a
divergence in the specific heat in the limit  $U \to U_0$. This is
clearly observed in Fig. \ref{fig21}b.


Other important quantities, useful for studying the critical
behavior of the system, are the charge and spin susceptibilities.
In fact, anomalies in their behaviors clearly signal the onset of
a CO state. The charge susceptibility can be defined separately
for the two sublattices:
\begin{equation*}
\chi _c^A =\frac{\partial n_A}{\partial \mu },\quad \quad \quad
\chi _c^B =\frac{\partial n_B}{\partial \mu },
\end{equation*}
where the total charge susceptibility is given by $ \chi _c
=\left(\chi _c^A +\chi _c^B \right)/2$.
\begin{figure}[t]
 \centering
 \subfigure[]
   {\includegraphics[scale=0.5]{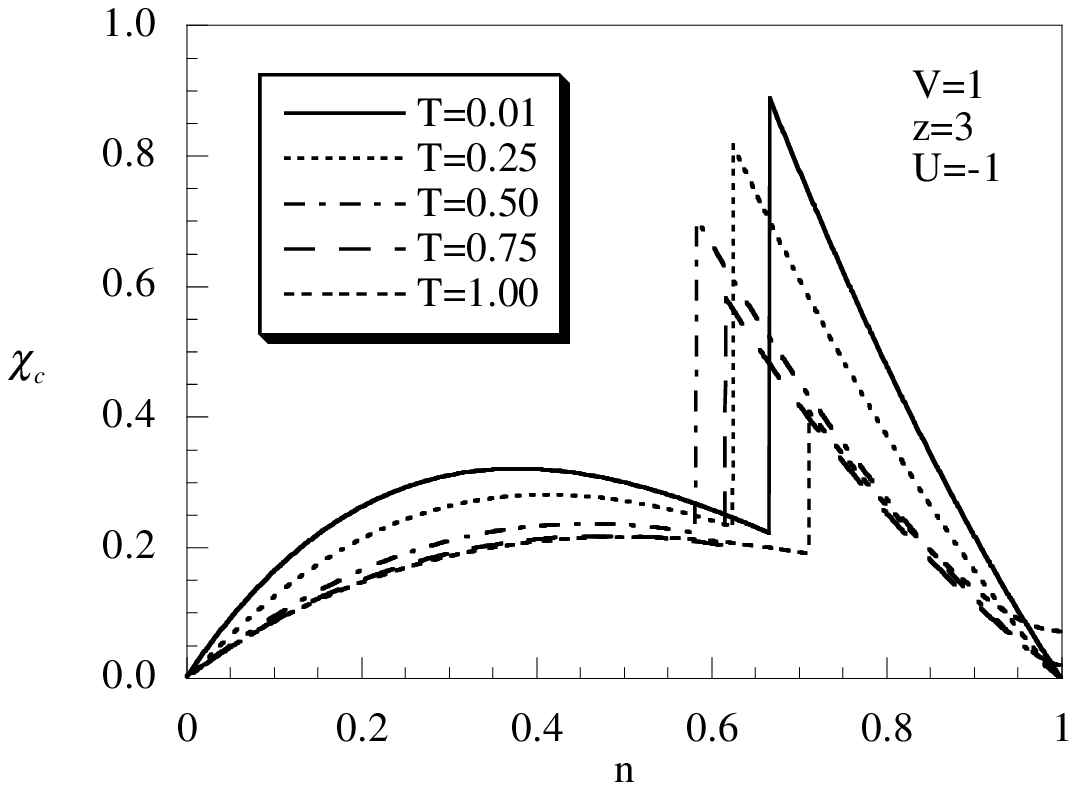}}
   \subfigure[]
   {\includegraphics[scale=0.5]{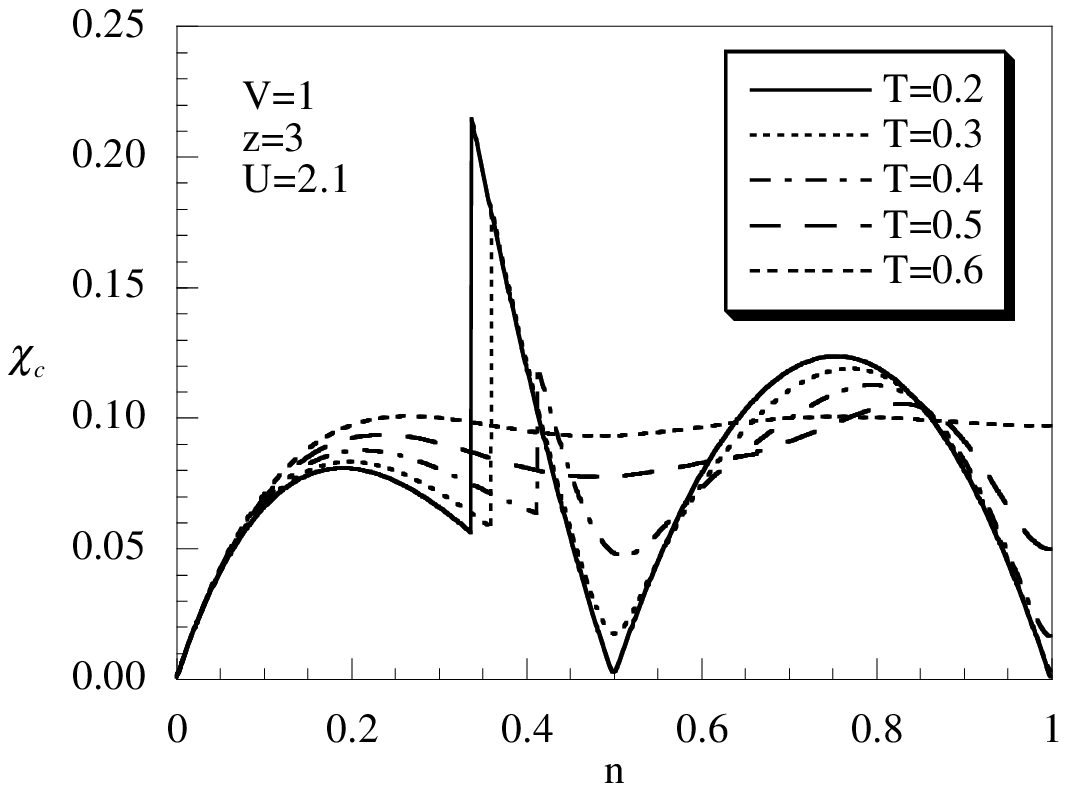}}
\caption{\label{fig72}The charge susceptibility as a function of
the particle density for $z=3$, $V=1$, different values of the
temperatures and (a) $U=-1$; (b) $U=2.1$.}
 \end{figure}
By studying separately $\chi _c^A$ and $\chi
_c^B$, it is manifest their different contribution  to the total
charge susceptibility. By varying
the particle density at fixed (low) temperature, one finds a
critical value of the particle density ($n_c$) at which the
susceptibilities $\chi _c^A$ and $\chi _c^{B}$ show a
discontinuity: below $n_c$, one finds $\chi _c^A =\chi _c^{B} =\chi_c$,
whereas above $n_c$ one observes the depletion of sublattice $A$
and the corresponding accumulation of the particles in the
sublattice $B$. For $n>n_c$, the susceptibility $\chi_c^A $
becomes negative, since $n_A$ is a decreasing function of $n$. As
a result, the system is in a CO phase with ``almost  empty"
shells separated by filled shells. Of course, at half filling and
in the limit $T \to 0$, one finds $n_A=0$ and $n_B=2$ and the
susceptibility vanishes. In Fig. \ref{fig72}, we plot the charge
susceptibility as a function of the particle density for a given
value of $U$ and several values of the temperature. At low
temperatures, $\chi_c$ shows a quasi-one or -two lobe structure,
depending on the value of the on-site potential. For $z=2$, i.e.,  a
1D chain, the lobes have a regular shape \cite{mancini08_PRE}. For
$z \ge 3$, the curvature of the lobes is different above and below
$n_c$ and the observed discontinuity signals the transition to a
phase where translational invariance is broken. The vanishing of
$\chi_c$ for $T \to 0$ at quarter and half filling indicates that
a full CO state is established.
In Figs. \ref{fig74}a-b, we plot $\chi _c$  as a function of the
temperature, for $U=V$ and $U=2.1V$, and for different values of
the filling ($n=0.25$, 0.5, 0.75, 1).
\begin{figure}[t]
 \centering
 \subfigure[]
   {\includegraphics[scale=0.5]{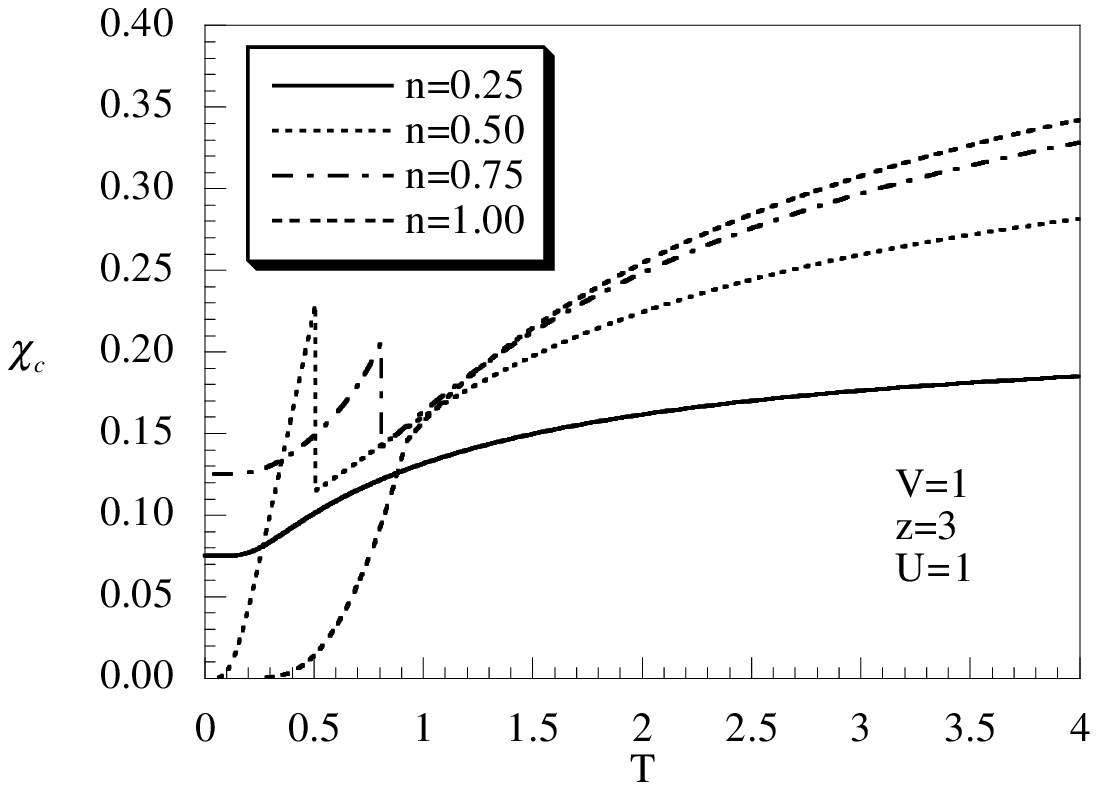}}
 \subfigure[]
   {\includegraphics[scale=0.5]{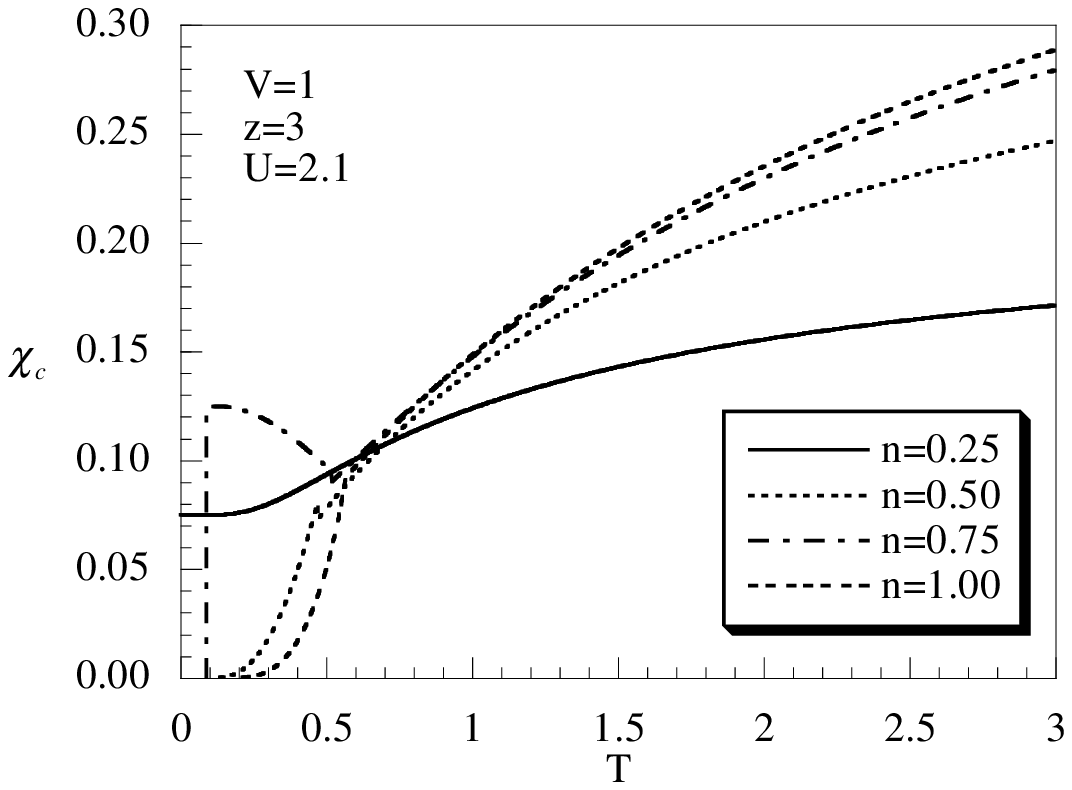}}
\caption{\label{fig74}  The charge susceptibility $\chi_c$ as a
function of the temperature for $z=3$, $V=1$, $n=0.25$, 0.5, 0.75,
1 and (a) $U=1$, (b) $U=2.1$.}
 \end{figure}
For $n=0.25$ there is no charge ordering and one finds $\chi _c^A
=\chi _c^B =\chi _c$. At $n=0.5$ and $n=0.75$, the system is in a
CO phase characterized by a different filling of the shells for
$T<T_c$: as a consequence, $\chi _c^A \ne \chi _c^B $, and
$\chi_c$ presents a discontinuity. At half filling, there is
charge ordering but the susceptibility is the same in the two
sublattices, also for $T<T_c$. The limit $T \to 0$ of the charge
susceptibility dramatically depends on the value of the particle
density: $\chi_c$ is finite for values of the particle density
corresponding to translational invariant states, whereas it
decreases with $T$ and vanishes for fillings corresponding to CO
states characterized by alternating empty and filled shells. In
the limit of high temperatures, the charge susceptibility tends to
a constant value which does not depend on $U$ and $z$ but only on
$n$ according to the law \cite{mancini08_PRE}
\begin{equation}
\lim_{T \to \infty } \chi_c =\alpha(n),
 \label{EHM_36}
\end{equation}
where $\alpha(n)=n(2 - n)/2$.

\begin{figure}[t]
 \centering
 \subfigure[]
   {\includegraphics[scale=0.5]{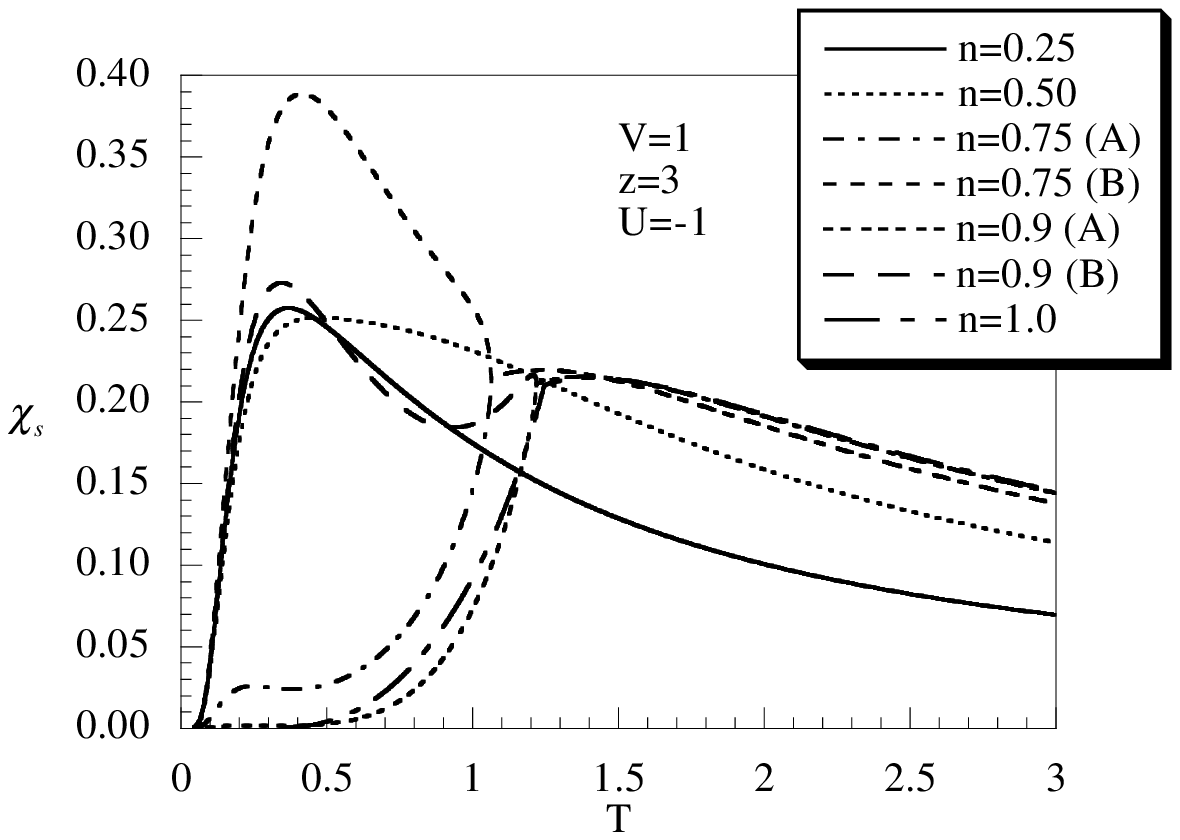}}
 \subfigure[]
   {\includegraphics[scale=0.5]{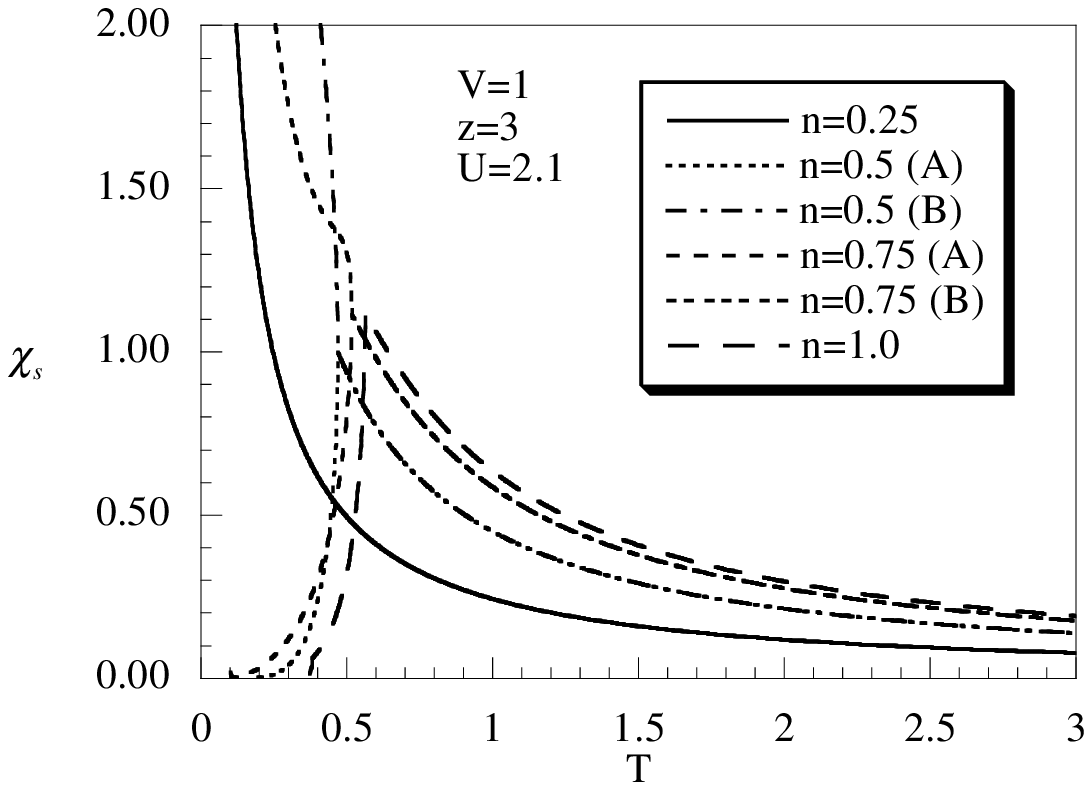}}
\caption{\label{fig76}(a) The spin susceptibilities  $\chi_s$,
$\chi_s^A$ and $\chi_s^B$ as a function of the temperature for
$V=1$, $n=0.25$, 0.5, 0.75, 1 and
 for (a) $U=-1$ and (b) $U=2.1$.}
 \end{figure}

The spin magnetic susceptibility $\chi_s$ in zero field, given
in Eq. \eqref{spin_susc}, can also be defined separately for the
two sublattices.
In Figs. \ref{fig76}a-b we plot the spin susceptibilities
$\chi_s$, $\chi_s^A$ and $\chi_s^B$ as functions of the
temperature for two representative values of $U$ ($U=-V$ and
$U=2.1V$, respectively) and for different values of the filling.
For $U=-V$, the spin susceptibility vanishes at zero temperature
for all values of the filling: all the electrons are paired and no
alignment of the spin is possible. By increasing $T$, the thermal
excitations break some of the doublons and a small magnetic field
may induce a finite magnetization: $\chi_s$ augments by increasing
$T$ up to a maximum, which might be different for the two
sublattices, then decreases. For $n=0.25$ and $n=0.5$ the system
is not charge ordered and $\chi_s$ has the same value in the two
sublattices. Conversely, for $n=0.75$ and $n=0.9$ the system is
charge ordered when $T<T_c$: $\chi_s$ assumes different values in
the two sublattices.

For $U=2.1V$, when the system is not charge ordered,
$\chi_s^A=\chi_s^B=\chi_s$ diverges for $T \to 0$, whereas for CO
states $\chi_s^A$ vanishes and $\chi_s^B$ diverges in the same
limit. At low temperatures, only the electrons belonging to
sublattice $A$ are paired and, as a consequence, $\chi_s^A= 0$. At
half filling, for both $U=-V$ and $U=2.1V$, even in the presence
of charge ordering, $\chi_s$ has the same value in the two
sublattices $A$ and $B$. It is easy to check that, for high
temperatures, the spin susceptibility decreases with the Curie law:
$\lim_{T \to \infty } \chi_s = \alpha(n)/T$ in the entire ($U,n$)
plane, where $\alpha(n)$ is the same  $z$ and $U$ independent
function appearing in Eq. \eqref{EHM_36} \cite{mancini08_PRE}. As
a consequence, in the limit of high temperatures, the ratio
$\chi_c/ \chi_s$ is an universal function of $T$, namely: $\lim_{T
\to \infty } \left( \chi_c / \chi_s \right) = T $.

\begin{figure}[t]
 \centering
 \subfigure[]
   {\includegraphics[scale=0.5]{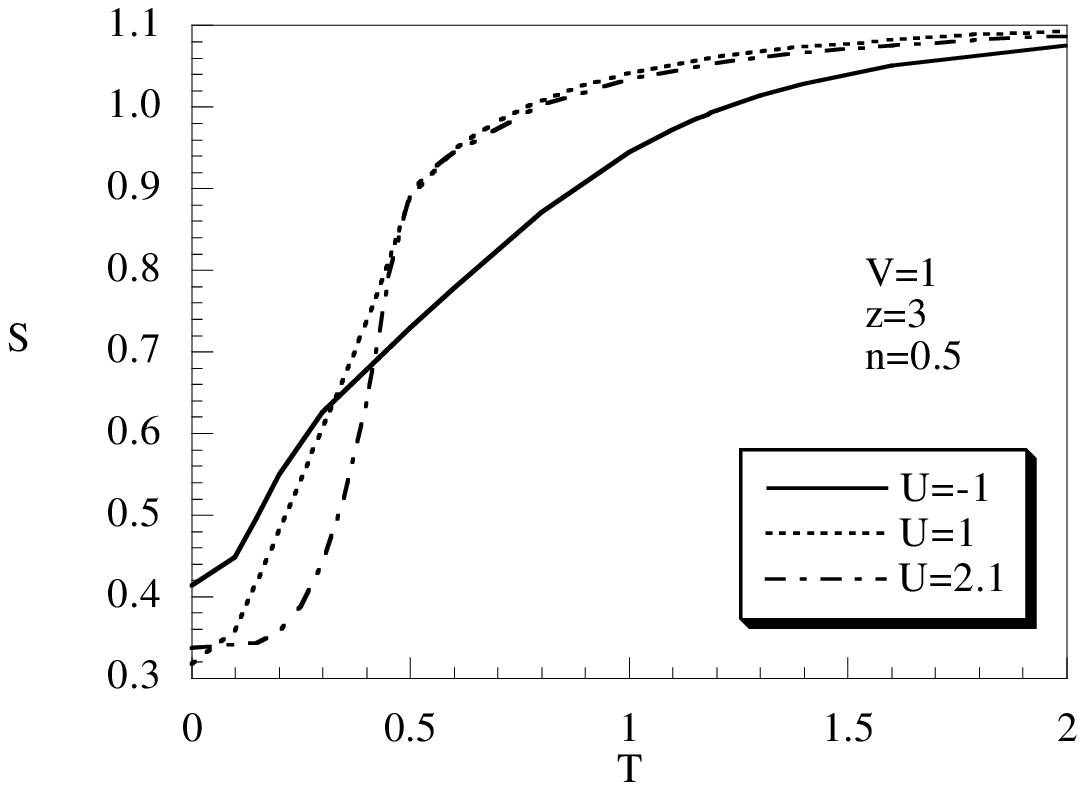}}
 \subfigure[]
   {\includegraphics[scale=0.5]{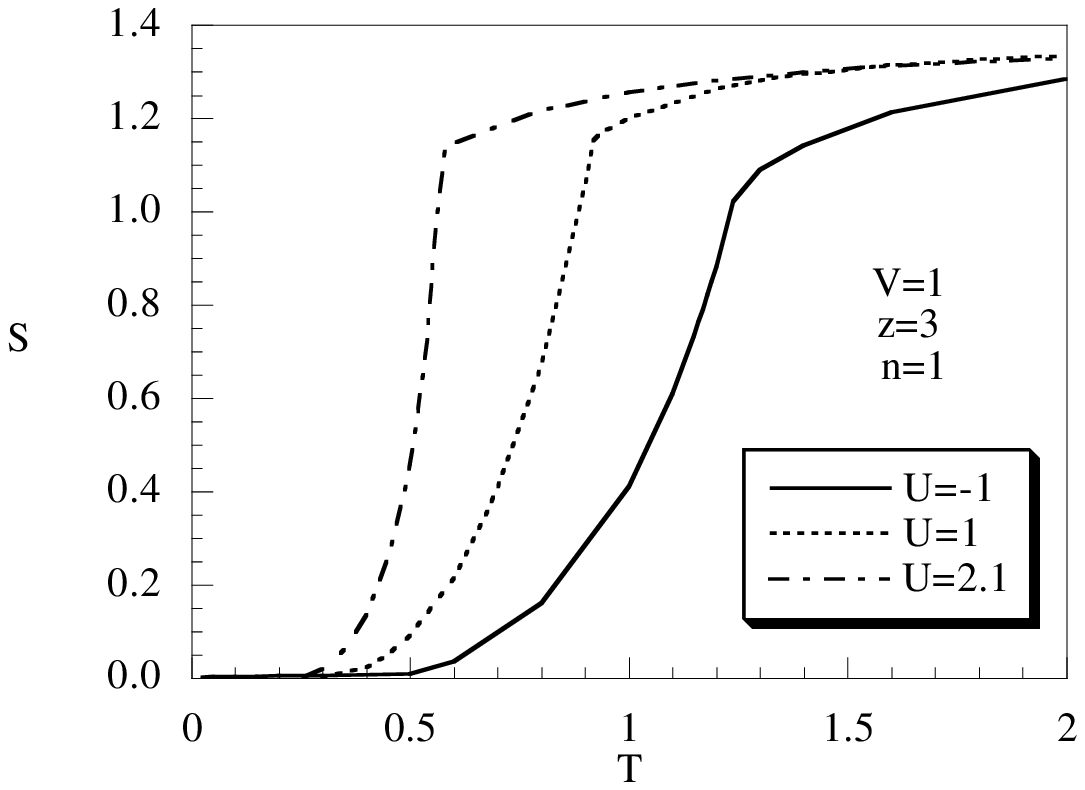}}
 \caption{\label{fig143} The entropy $S$ as a function of the temperature
 for $z=3$, $V=1$, $U=-1$, 1, 2.1 and (a) $n=0.5$, (b) $n=1$.   }
 \end{figure}
\begin{figure}[t]
 \centering
 \includegraphics[scale=0.57]{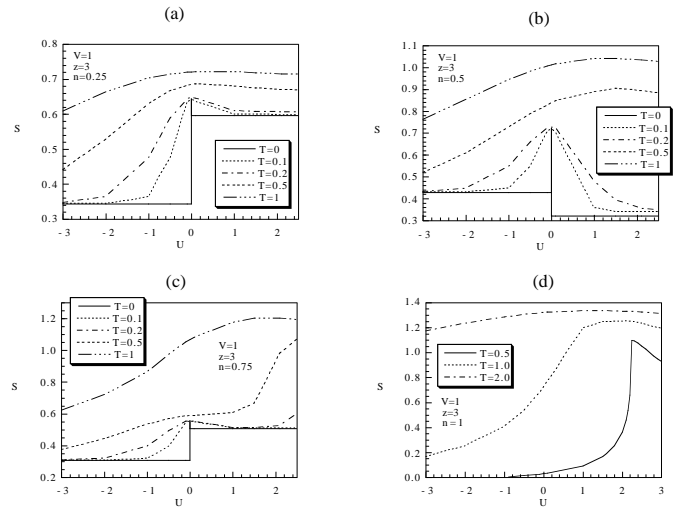}
\caption{\label{fig146} The entropy $S$ as a function of $U$ for
$z=3$, $V=1$, for temperatures varying in the interval (0,1) and
(a): $n=0.25$, (b): $n=0.5$, (c): $n=0.75$, (d): $n=1$.}
 \end{figure}

To conclude this section, we report some results obtained for the
entropy as a function of $T$, and $U$, showing that also this quantity is
a good indicator of the onset of a CO state. The standard way to compute
the entropy is by integrating via the integral of the specific heat:
\begin{equation*}
S(T)=S(0)+\int_0^T \frac{C(T)}{T} \, dT.
\end{equation*}
However, as extensively commented on in Ref. \cite{mancini08_PRE},
this expression can not be easily handled since it requires the
calculation of $S(0)$, which is  generally not an easy task. A more
convenient formula is given by:
\[
 S(n,T,U)=-\int_0^n \frac{\partial \mu (n',T,U)}{\partial T} \, dn'.
\]
In our framework of calculations, we can readily deal with this expression
since it requires only the knowledge of the chemical
potential. In Figs. \ref{fig143}a and \ref{fig143}b we plot the
entropy as a function of the temperature for $n=0.5$ and $n=1$,
respectively.
One can notice an abrupt change of the entropy curves  when
the critical temperature is reached. One may also notice that, in the limit $T \to 0$,
$S(0)$ is finite for $n=0.5$ due to the
degeneracy of the ground state energy. On the other hand, $S(0)$ is zero at half
filling because there is no degeneracy of the ground state.
The $U$ dependence of the entropy is rather dramatic in the
neighborhood of the values at which a zero temperature transition
occurs, as it is evident from Fig. \ref{fig146}. At low
temperatures, the entropy presents a step-like behavior and
becomes rather insensitive to variations in $U$ for sufficiently
large on-site repulsive and attractive interactions. For
$0<n<1/z$, one observes an increase of the entropy at $U=0$ in the
limit $T \to 0$: even if there is no transition, there is a
change of the distribution of the electrons from a configuration
characterized by doubly occupied sites to one with singly occupied
sites, which is less ordered. When $1/z<n <2/z$, the entropy
presents a discontinuity  around $U=0$ (due to the phase
transition) which becomes more pronounced as the temperature
decreases. In the region $2/z<n<1$, at low temperatures, one finds
a rather sharp increase of the entropy near $U_0(n)$, where the
system undergoes a transition from a CO phase to a translational
invariant state, less ordered.
%

\section{concluding remarks}
\label{sec_V}

Statistical models on the Bethe lattice are of considerable
interest since they admit a direct analytical approach for a
number of problems that may be otherwise intractable on Euclidean
lattices. In this paper, we have evidenced how the use of the
Green's function and equations of motion formalism leads to the
exact solution of the extended Hubbard model on the Bethe lattice
in the narrow-band limit. We provided a comprehensive and
systematic analysis of the model by considering relevant
thermodynamic quantities in the whole space of the parameters $n$,
$T$, $U$ and $V$ and we obtained the finite temperature phase
diagram, for both attractive and repulsive on-site and intersite
interactions.

The phase diagram dramatically depends on the sign of the
nearest-neighbor interaction. In the attractive case, there is a
critical temperature $T_c$, located by the divergence of the
charge susceptibility, at which there is a transition from a
thermodynamically stable to an unstable phase; the latter is
characterized by phase separation. The surface separating the two
phases has a rounded vault in the 3D ($n$, $U$, $T$) space, which
splits in two for $U>U_c(z)$. Below $T_c$, the same critical value
of the on-site potential separates the two possible configurations
of phase separated states, namely clusters of singly or doubly
occupied sites. For repulsive nearest-neighbor interactions, the
phase diagram has a richer structure:  we found a transition
temperature below which translational invariance is broken. The
Bethe lattice effectively splits in two sublattices with different
thermodynamic properties. As a result, a charge ordered phase,
characterized by a different distribution of the electrons in
alternating shells, is established for $n>1/z$. The CO phase is
energetically favored as demonstrated by the study of the
Helmholtz free energy. By investigating the behavior of several
thermodynamic quantities, we attained a full comprehension of the
phase diagram. The study of the particle density and of the double
occupancy is particularly enlightening to unveil the distribution
of the particles on the sites of the Bethe lattice. The specific
heat exhibits not only high and low temperature peaks due to
charge excitations induced by the Hubbard operators $\psi^{(\xi)}$
and $\psi^{(\eta)}$, but also peaks  due  to the phase transition
from the homogeneous phase to the CO state (or viceversa). By
studying separately the contribution to the charge and spin
susceptibilities coming from the two sublattices, the onset of a
CO phase is also signalled by the separation of the values of the
sublattices quantities $\chi_{c,s}^A$ and $\chi_{c,s}^B$ . The
different population of the two sublattices in the CO phase is
reflected by the divergence or the vanishing, in the limit $T \to
0$, of the spin susceptibilities $\chi_s^A$ and $\chi_s^B$. If the
electrons are paired, no alignment of the spin is possible
($\chi_s^A \to 0$), whereas single occupation of the sites leads
to an alignment ($\chi_s^B \to \infty$) even when the magnetic
field is turned off.

\section*{Acknowledgements}
We thank A. Naddeo for her contribution to the initial stages of this work
and A. Avella for stimulating discussions and a careful
reading of the manuscript.
\appendix

\section{T\MakeLowercase{he extended} H\MakeLowercase{ubbard model on the} B\MakeLowercase{ethe Lattice in the
presence of an external magnetic field}} \label{app_A}


In the presence of an external magnetic field $h$,  the Hamiltonian of the AL-EHM reads:
\begin{equation}
\label{eq821}
\begin{split}
H&=-\mu \sum_i n(i)+U\sum_i
D(i)\\
&+\frac{1}{2}\sum_{i\ne j} V_{ij} n(i)n(j)-h\sum_i n_3 (i),
\end{split}
\end{equation}
where $n_3 (i)$ is the third component of the spin density
operator
\begin{equation}
\label{eq822} n_3 (i)=n_\uparrow (i)-n_\downarrow (i)=c_\uparrow
^\dag (i)c_\uparrow (i)-c_\downarrow ^\dag (i)c_\downarrow (i).
\end{equation}
The formulation given in Section \ref{sec_II} for the case of a
Bethe lattice must be generalized in order to take into account
the breaking of rotational invariance in the spin space: one has
to distinguish the two components - in the spinorial notation  -
of the fermionic fields. By exploiting the decomposition
\eqref{EHM_17}, it is not difficult to show that the presence of
the magnetic field affects only  $H_0$. In this representation the
equations of motion of the Hubbard operators become
\begin{equation}
\begin{split}
\left[ \xi_{\sigma} (i),H_0 \right] &= - \left( \mu +\sigma h  \right) \xi_{\sigma} (i) ,\\
\left[ \eta_{\sigma} (i),H_0 \right] &= -\left(\mu +\sigma h-U
\right)\eta_{\sigma} (i),
 \end{split}
 \end{equation}
and  Eqs. \eqref{eq_nD_1} are modified as:
\begin{equation}
\label{eq8217}
\begin{split}
\langle n_\uparrow (i)\rangle _0 &=\frac{e^{\beta (\mu +2h)}+e^{\beta (2\mu
+h-U)}}{e^{\beta h}+e^{\beta \mu }+e^{\beta (\mu +2h)}+e^{\beta
(2\mu
+h-U)}} ,\\
\langle n_\downarrow (i)\rangle _0 &=\frac{e^{\beta \mu }+e^{\beta (2\mu
+h-U)}}{e^{\beta h}+e^{\beta \mu }+e^{\beta (\mu +2h)}+e^{\beta
(2\mu
+h-U)}} ,\\
 \langle D(i)\rangle _0 &=\frac{e^{\beta (2\mu +h-U)}}{e^{\beta h}+e^{\beta \mu
}+e^{\beta (\mu +2h)}+e^{\beta (2\mu +h-U)}} ,\\
 \langle n_3 (i)\rangle _0 &=\frac{e^{\beta \mu
}(e^{2\beta h}-1)}{e^{\beta h}+e^{\beta \mu }+e^{\beta (\mu
+2h)}+e^{\beta
(2\mu +h-U)}} .
 \end{split}
\end{equation}
\vspace{0.5cm} All the rest of formulation developed in Sections
\ref{sec_II}, \ref{sec_III} and \ref{sec_IV} follows easily. It is
worth noticing that the magnetization takes the simple expression:
\begin{equation}
\langle n_3 (i)\rangle =\tanh (\beta h)[\langle n(i)
 \rangle -2 \langle D(i)\rangle].
\end{equation}

\end{document}